\documentclass{aa}  
\usepackage{graphicx}
\usepackage{rotating,booktabs}
\usepackage[verbose]{placeins}
\usepackage{pdflscape}
\usepackage{adjustbox}
\usepackage{tikz}
\usepackage[english]{babel}

\usepackage{txfonts}
\usepackage{natbib}
\usepackage{longtable}

\newcommand{\lya}{\rm Ly$\alpha~$}

\newcommand{\lumcgs}{erg~s$^{-1}$}

\newcommand{\lbol}{\rm L_{ bol}}

\begin{document} 

   \title{The WISSH quasars project}
   \subtitle{VII. Outflows and metals in the circumgalactic medium around the hyper-luminous $z\sim$3.6 quasar J1538+08\thanks{Based on data obtained with the European Southern Observatory Very Large Telescope, Paranal, Chile, under Program 099.A-0316(A)}}

   \titlerunning{The WISSH quasar project VII}
   
   \author{A. Travascio
          \inst{1,2,3}
          \and
          L. Zappacosta
          \inst{1}
          \and
          S. Cantalupo
          \inst{4}
          \and 
          E. Piconcelli
          \inst{1}
          \and 
          F. Arrigoni Battaia
          \inst{5,6}
          \and 
          M. Ginolfi
          \inst{7}
          \and 
          M. Bischetti
          \inst{1}
          \and 
          G. Vietri
          \inst{8}
          \and 
          A. Bongiorno
          \inst{1}
          \and 
          V. D'Odorico
          \inst{9,10}
          \and 
          F. Duras
          \inst{11}
          \and 
          C. Feruglio
          \inst{9}
          \and 
          C. Vignali
          \inst{12,13}
          \and 
          F. Fiore
          \inst{9}
          }

   \institute{INAF–Osservatorio Astronomico di Roma,Via Frascati 33,00078 Monteporzio Catone, Italy \\ email: \texttt{andrea.travascio@inaf.it} \and 
   Department of Physics, University of Rome "Tor Vergata", Via della Ricerca Scientifica 1, I-00133 Rome, Italy \and
   Università degli Studi di Roma "Sapienza", Piazzale Aldo Moro 5, 00185 Roma, Italy \and 
   Department of Physics, ETH Zurich, Wolfgang-Pauli-Strasse 27, 8093 Zurich, Switzerland \and
   European Southern Observatory, Karl-Schwarzschild-Str. 2, D-85748, Garching bei Munchen, Germany \and
   Max-Planck-Institut fur Astrophysik, Karl-Schwarzschild-Str 1, D-85748 Garching, Germany  \and 
   Observatoire de Genève, Université de Genève, 51 Ch. des Maillettes, 1290 Versoix, Switzerland \and 
   INAF, Istituto di Astrofisica Spaziale e Fisica Cosmica - Milano, via A. Corti 12, I-20133, Milano, Italy \and 
   INAF Osservatorio Astronomico di Trieste, Via G.B. Tiepolo, 11, Trieste, I-34143, Italy \and 
   Scuola Normale Superiore, Piazza dei Cavalieri 7, I-56126 Pisa, Italy \and 
   Dipartimento di Matematica e Fisica, Università Roma Tre, via della Vasca Navale 84, 00146 Roma, Italy \and
   Dipartimento di Fisica e Astronomia, Università di Bologna, Via Piero Gobetti 93/2, I-40129 Bologna, Italy \and 
   INAF–Osservatorio di Astrofisica e Scienza dello Spazio di Bologna, Via Piero Gobetti 93/3, I-40129 Bologna, Italy}
             


  \abstract
   {During the last years, Ly$\alpha$ nebulae have been routinely detected around high redshift, radio$-$quiet quasars thanks to the advent of the highly sensitive integral field spectrographs. Constraining the physical properties of the Ly$\alpha$ nebulae is crucial for a full understanding of the circum-galactic medium (CGM). The CGM acts both as repository for intergalactic/galactic baryons and a venue of feeding and feedback processes. The most luminous quasars are privileged test-beds to study these processes, given their large ionizing fluxes and dense CGM environments in which they are expected to be embedded.}
   {We aim at characterizing the rest-frame ultra-violet (UV) emission lines in the CGM around a hyper-luminous, broad emission line, radio-quiet quasar at z$\sim$3.6, that exhibits a powerful outflows at both nuclear and host galaxy scales.}
   {We analyse VLT/MUSE observations of the quasar J1538+08 ($\rm L_{bol}=6 \times 10^{47}~erg~s^{-1}$) and perform a search for extended UV emission lines to characterize its morphology, emissivity, kinematics and metal content. }
   {We report the discovery of a very luminous  ($\sim2\times10^{44}\rm ~erg~s^{-1}~cm^{-2}$), giant (150~kpc) Ly$\alpha$ nebula and a likely associated extended (75 kpc) CIV nebula. The Ly$\alpha$ nebula emission exhibits moderate blueshift ($\sim$440$\rm ~km~s^{-1}$) compared with the quasar systemic redshift and a large average velocity dispersion ($\bar{\sigma _{\rm v}} \sim700\rm ~km~s^{-1}$) across the nebula, while the CIV nebula shows average velocity dispersion $\rm \bar{\sigma _{\rm v}} \sim350\rm ~km~s^{-1}$.
   The Ly$\alpha$ line profile exhibits a significant asymmetry towards negative velocity values at 20$-$30 kpc south of the quasar and is well parametrised by two Gaussian components: a narrow ($\sigma\sim470\rm~km~s^{-1}$) systemic one plus a broad ($\sigma\sim1200\rm~km~s^{-1}$), blueshifted ($\sim1500\rm~km~s^{-1}$) one.
   }
   {
   Our analysis of the MUSE observation of J1538+08 reveals metal-enriched CGM around this hyper-luminous quasar. Furthermore, our detection of blueshifted emission in the emission profile of the Ly$\alpha$ nebula suggests that powerful nuclear outflows can propagate through the CGM over tens of kpc.}


   \maketitle
%

\section{Introduction}
In the past few decades, the investigation of extended (tens of kpc-scales) $\rm Ly\alpha$-emitting nebulae surrounding active galaxies have become more and more intense (see \citealt{Cantalupo17} and references therein). These are a promising repository for the census of baryonic matter and metals and a privileged environment to study the feeding and feedback processes.
The first Circum-Galactic Medium (CGM) Emission Nebulae detected in Ly$\alpha$ (Ly$\alpha$-CEN hereafter) were mostly observed around high redshift radio galaxies with sizes exceeding 100 kpc, through Narrow Band (NB) -imaging and slit-spectroscopy \citep[e.g.][]{Heckman91,vanOjik97,Villar06,Villar07,Humphrey07,Humphrey13}. Only a minor fraction ($\sim$10$\%$) of them was reported around radio-quiet quasars \citep[RQQs; e.g.][]{Weidinger05,Christensen06,Battaia16}.
In rare cases, RQQs have been found lying in overdense AGN environments and embedded in Enormous Ly$\alpha$ Nebulae (ELANe) with projected sizes exceeding $\sim$200~kpc and surface brightnesses (SB) >$\rm 10^{-17} erg~s^{-1}~cm^{-2}~arcsec^{-2}$ for hundreds of kpc \citep{Cantalupo14,Hennawi15,Cai17}. 
Recent sensitive integral field spectrographs, like VLT/MUSE \citep[Multi Unit Spectroscopic Explorer;][]{Bacon10} and Keck/KCWI (Keck Cosmic Web Imager), have recently allowed the detection of extended Ly$\alpha$ emission around z=2-5 quasars at a depth never explored before and provided a 3D view of Ly$\alpha$-CEN.
\cite{Borisova16} and \cite{Battaia19} (hereafter B16 and AB19, respectively) performed MUSE surveys of optically-bright RQQs at 3<z<4 and \cite{Cai19} used the KCWI to study the CGM surrounding RQQs at z$\approx$2, reporting detection rates of nearly 100$\%$ for Ly$\alpha$-CEN.

Detailed morphological and kinematic studies of Ly$\alpha$-CEN reveal that they have large ($\sim$100 kpc diameter) and mostly symmetrical  structures with the bulk of their emission concentrated within tens of kpc \citep[B16; AB19; ][]{Ginolfi18,Lusso19}. Their average SB profiles have been parameterized with exponential (AB19) or power-law (B16) profiles with no clear dependence on radio-loudness (AB19). Although with a few exceptions \citep[e.g.][]{Weidinger04,Weidinger05,Cai17b,Battaia18b}, Ly$\alpha$-CEN do not show any clear kinematic pattern (e.g., ascribed to ordered gas rotation or inflows/outflows) and exhibit velocity dispersions $\rm \sigma _v \lesssim~400~km~s^{-1}$. The latter are consistent with the expected kinematics of their dark matter halo gravitational potential. However, higher $\rm \sigma _v$ values have been reported for CEN around some RQQs and radio-loud sources \citep[B16; ][]{Ginolfi18}. For the latter, the high $\rm \sigma _v$ is likely due to the jets mechanical interaction with the Ly$\alpha$-CEN \citep{vanOjik97,Villar06,Silva18}.

To date, at least three mechanisms have been proposed to explain extended Ly$\alpha$ emission of cold ($\rm T \sim10^4 ~K$) gas.
The main one is fluorescence, which was predicted for the first time by \cite{Hogan87}, whereby the gas emits Ly$\alpha$ photons by recombination when photo-ionized by one or more UV sources \citep{Cantalupo05}. The second is associated with AGN Ly$\alpha$ photons, which can be resonantly scattered by the neutral hydrogen clouds \citep{Moller98,Cantalupo05}. Finally, shocks triggered by galactic outflows can power Ly$\alpha$ emission \citep{Taniguchi00,Battaia15b}.
The contribution of all these mechanisms is expected to be large in the CGM surrounding hyper-luminous ($\rm \lbol\gtrsim10^{47} erg~s^{-1}$) quasars. Indeed, the latter are the most luminous UV emitters \citep{WISSHII} and sits on large potential wells, likely suggesting the presence of relatively high-density CGM and significant over-density of companions \citep{Bischetti18}.
Furthermore, luminous quasars are also predicted and observed to launch the most powerful winds out to host-galaxy scales \citep{Faucher12,Fiore17,Menci19}. Hence, we can reasonably expect them to affect the kinematics of the CGM and its metal content.

In this paper we report VLT/MUSE observations of SDSS~J153830.55+085517.0 (hereafter J1538+08), a hyper-luminous ($\rm L_{bol} \approx6\times10^{47} erg~s^{-1}$) quasar at $z _{\rm QSO} = 3.567_{-0.002} ^{+0.003}$ \citep[based on H$\beta$;][]{Vietri18}. This source belongs to the WISSH sample \citep{WISSHI}, 
i.e. the optically/mid-infrared selected most luminous quasars at $z$=2-4.
The WISSH quasars are characterized by pervasive powerful winds from nuclear \citep{Vietri18,Bruni19} out to kpc scales \citep{WISSHI} and, hence, are ideal laboratories for a detailed study of the kinematics and metal content at CGM scales.

The paper is organized as follows. Section~\ref{sec:data} and \ref{sec:detneb} describe the reduction of the MUSE data and the methodology for CEN detection, respectively. Section~\ref{detectGLAN} presents the results of our analysis along with the physical properties of the CEN revealed around J1538+08. Section~\ref{sec:discussion} is devoted to the discussion of the Ly$\alpha$-CEN properties, the presence of metals (i.e. CIV-CEN) and the evidence of an outflow in the CGM traced by the blueshifted Ly$\alpha$ emission. Finally, Section~\ref{conclusions} reports our conclusions. 
Throughout the paper we adopt a cosmology with $\Omega_\Lambda=0.68$ and $H_0=67.4 \,\rm  km\, s^{-1} Mpc^{-1}$ \citep{Plank18}, for which 1~arcsec corresponds to $\sim$7.4~kpc at the quasar redshift. All the errors are quoted at 1$\sigma$ significance and all flux-weighted quantities are computed in regions with signal to noise $\geq$3, unless otherwise stated.

\section{Data Reduction} \label{sec:data}

J1538+08 was observed with MUSE on July 26 2017, as part of the ESO program ID 099.A-0316(A) (PI F. Fiore). 
The observation consists of 4 exposures of 1020 sec each, for a total integration time of 1 h and 8 minutes. Each exposure was rotated by 90 degrees  with the addition of a small dithering. The average seeing was $\approx 0.9$~arcsec.  
The data reduction was performed by using the ESO MUSE pipeline \citep[EsoRex v. 3.12.3;][]{Weilbacher14} and the code \texttt{CubExtractor} \citep[CubEx v. 1.8; see][]{Cantalupo19}. 
We followed a procedure similar to that reported in B16, which has been described in more detail by \cite{Cantalupo19}. We briefly summarize the main steps in the following:
\begin{itemize}
    \item For each exposure we apply bias subtraction and correct for flat-fielding, twilight  and illumination, using the standard ESO MUSE pipeline. Finally, we apply a wavelength, geometry and astrometric calibration. 
    
    
  \item we use CubeFix (a tool from the CubExtractor Package) to apply a more accurate flat-fielding correction in each datacube by using the continuum and the emission lines of the sky as calibrators. This procedure consists in deriving the correction factors to be applied to each integral field unit (IFU) and its elements to have coherent values of the sky on the whole field of view (FOV). This allows us to perform a wavelength and flux dependent correction. The continuum of the sources need to be iteratively masked to minimize the self-calibration errors;  
    


  \item We perform a correction of the sky line spread function with the \texttt{CubeSharp} procedure, which adopts a flux-conserving sky-subtraction method \citep[see][]{Cantalupo19}.

 \item Then \texttt{CubEx} is used to obtain more refined science datacubes and the associated variances. Notice that, the standard ESO pipeline can underestimate the propagated variance \citep{Bacon17}. This variance is therefore propagated during the execution of \texttt{CubExtractor} packages and rescaled by a constant factor to match the empirical spatial variance estimated from the cube at each wavelength layer \citep[see][for details]{Borisova16}. The final datacube is a median stack (with 3$\sigma$-clipping) of the single datacubes derived from each exposure.
 \end{itemize}

Finally, we notice that during the data reduction with the \texttt{CubExtractor} code,  we account for the presence of a saturated bright (Vega magnitude V $\simeq$ 9.7) star at the edge of the MUSE FOV contaminating large part of the field in each exposure. In the flat-fielding and sky-subtraction, we indeed mask the stellar emission and the bleed trails up to a distance of 16 arcsec from the quasar where this contribution starts to become negligible for our purposes (see Appendix.~\ref{Saturstar} for further details).

\section{CEN detection methodology} \label{sec:detneb}

We performed the identification and extraction of the CEN in the final datacube with the \texttt{CubExtractor} package.
We first removed the quasar Point Spread Function (PSF) with the \texttt{CubePSFSub} task, by masking the expected spectral regions having an extended emission in order to avoid any contamination in the PSF estimation. This recipe estimates the quasar PSF empirically at each wavelength layer from a pseudo-NB image. In our case, each pseudo-NB image was computed with a spectral width of $\pm150$ pixels, corresponding to $\pm188$ \AA. The estimated PSF was rescaled to its flux in each wavelength layer image and then subtracted. The PSF normalization was derived by assuming that the central pixels, within a region of $1'' \times 1''$, are quasar dominated.
Furthermore, we removed the continuum emission of any source within the MUSE FOV by means of a fast median-filtering approach using the \texttt{CubeBKGSub} task (see \citealt{Cantalupo19} for further details).

We searched for diffuse emission in the PSF- and continuum- subtracted datacube with the \texttt{CubEx} recipe. 
The latter allows us to detect and extract extended sources, by applying a connected labeling component algorithm. The key input parameters of this algorithm are  $N^{vox}_{min}$ and $SNR_{th}$, which define the minimum number of connected voxels and the Signal to Noise Ratio (SNR) threshold of the extended emission to be searched for, respectively. \texttt{CubEx} returns astrometric, photometric and spectroscopic information on the extended nebular emissions, and generates both a three-dimensional mask (3D-mask), which defines the datacube elements (i.e. voxels) belonging to the detected nebula, and a three dimensional SNR cube.
Once the CEN is detected and extracted, the \texttt{Cube2Im} task allows us to create, for the detected CEN, three data products by using the voxels of the PSF- and continuum- subtracted datacube defined in the 3D-mask:
(i) the SB map, i.e. an "optimally extracted image" derived by applying the 3D-mask to the datacube and by collapsing at each spaxel the contribution of the nebula voxels above the chosen $SNR_{th}$; (ii) a map of the velocity distribution obtained as the first moment of the flux distribution ; (iii) a map of $\sigma _{\rm v}$ which is derived as the second moment of the flux distribution. All these data products were smoothed with a Gaussian-kernel of $\sigma = 2$ pixels (0.4 arcsec).
We computed the SNR map from the pseudo-NB image obtained by collapsing the wavelength range (i.e. the layers) in which the extended emission was detected to have an estimation of the variance distribution of the SB map. Then we associated a SNR value to each pixel, where the noise is the average of the standard deviations derived in several background regions with size $1'' \times 1''$. The SNR map was finally smoothed with a Gaussian-kernel of $\sigma$= 2 pixels, to be consistent with the smoothing of the maps produced by \texttt{Cube2Im}.

\section{Results}\label{detectGLAN}

We searched for CEN around J1538+08 traced by typical CGM UV transitions, such as Ly$\alpha$ $\lambda$1215 \AA, NV$\lambda$1240 \AA, SiIV$\lambda$1397 \AA, CIV$\lambda$1549 \AA, HeII$\lambda$1641 \AA~ and CIII]$\lambda$1909 \AA. 
We detected significant nebular emission in Ly$\alpha$ and CIV. The properties of the detected CEN are described in the following sections.

\begin{figure}[t]
   \begin{center}
   \includegraphics[height=0.38\textheight,angle=0]{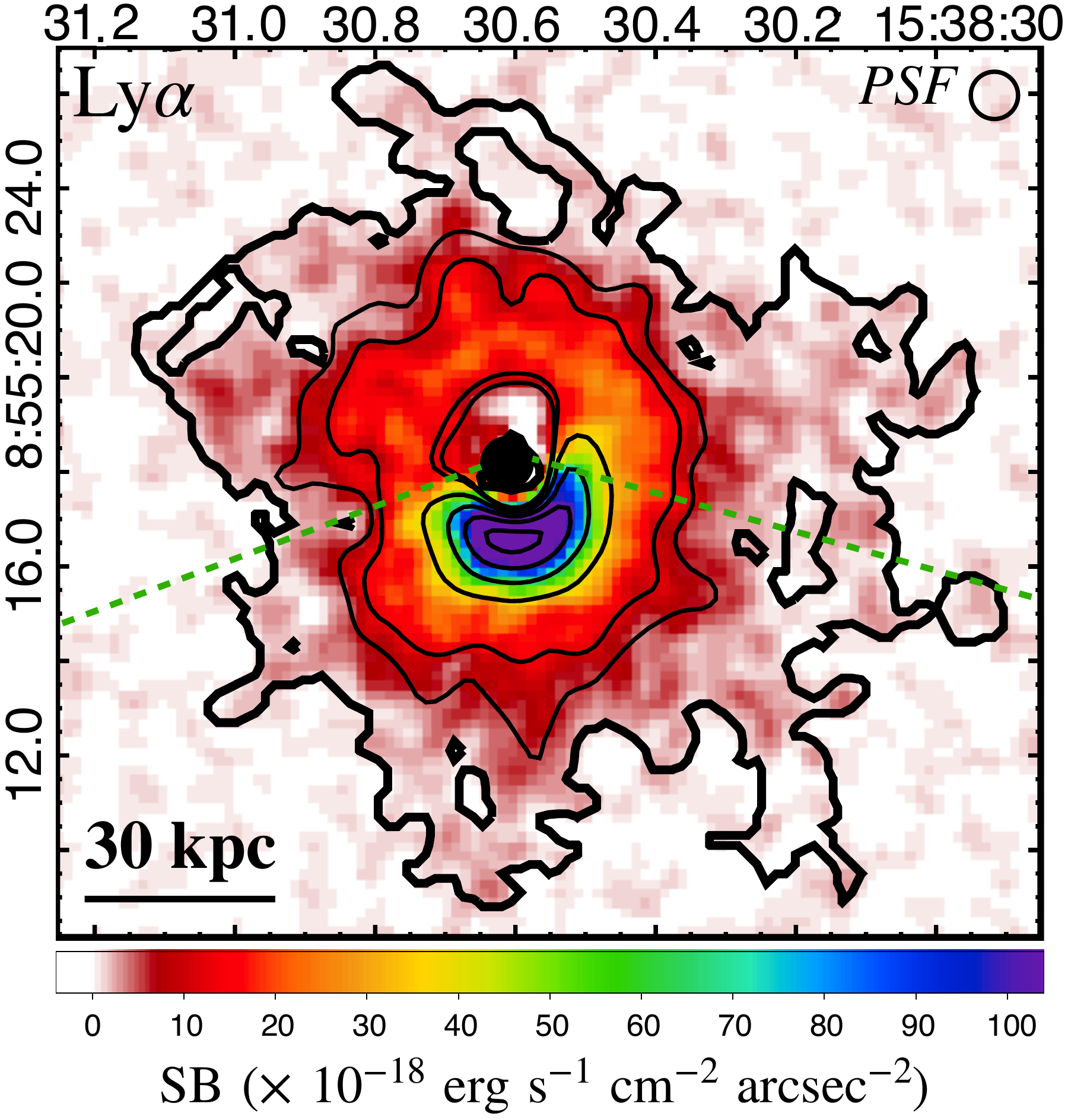}
   \caption{Optimally extracted SB map of the Ly$\alpha$-CEN. The quasar position is marked by a black dot. The thick black contour corresponds to the two-dimensional projection of the boundary of the CEN from the 3D-mask. It indicates a SB level of about $\rm 10^{-18}~erg~s^{-1}~cm^{-2}~arcsec^{-2}$. The thin black contours indicate SNR levels $3, 5, 15, 25, 35, 45$ extracted from the SNR map. The wavelength layer at the Ly$\alpha$ profile peak is used to report the background fluctuations in the image. On the top-right corner the circle indicates the size of the FWHM of the instrument PSF. The green dashed lines delimit two regions of the Ly$\alpha$-CEN (i.e. north and south from the quasar position) used for the extraction of the radial profile (see Fig.~\ref{figures:rpLya}). The low-SNR region close to the quasar with SB values around zero could be the result of either a random fluctuations or a real features of the distribution of the CGM gas emission.}
   \label{figure:SBLya}
   \end{center}
\end{figure}
\setlength{\tabcolsep}{2pt}
\begin{table}[!t] 
\begin{center}
\caption{Physical properties of the detected CEN}
\label{table:NebulaProp}
\begin{tabular}{lrrr}
\hline
\hline
                             &  Ly$\alpha$    &     CIV    \\[3pt]
\hline        
$\rm \lambda _{CEN}$ [\AA]$^{a}$         &  5543.8  $\pm$ 1.8 & 7064.7 $\pm$ 1.8 \\[3pt]
$z _{\rm CEN}$ $^{b}$  & 3.560 $\pm$ 0.002 &  3.559 $\pm$ 0.001  \\[3pt]
Size[kpc]$^{c}$ &  150 & 75 \\[3pt]
Spectral width [\AA]$^{d}$     & 95 &   42 \\
$\rm d_{CEN}^{peak}$ [kpc]$^{e}$  & 10.4$\pm$ 3.7$^{\star}$  & 15.1$\pm$ 3.7$^{\star}$ \\[3pt]   
$\rm d_{CEN}^{cen}$ [kpc]$^{f}$  & 5.7$\pm$ 3.7$^{\star}$ & 13.7$\pm$ 3.7$^{\star}$  \\[3pt]
$\alpha$ $^{g}$ & 0.72 & 0.62 \\[3pt]
$\phi$ [degree]$^{h}$  & 62 & 11 \\[3pt]
Flux [$\rm \times 10^{-16}~erg~s^{-1}~cm^{-2}$]$^{i}$ &      17.8$ \pm $0.1     &     0.77 $\pm$ 0.03  \\[3pt]
Luminosity [$\rm \times 10^{43}~erg~s^{-1}$]$^{l}$ & 20.55 $\pm$ 0.16 & 1.00 $\pm$ 0.03 \\[3pt]
$\bar{\sigma}_v$  [$\rm km~s^{-1}$]$^{m}$ &  770 $\pm$ 2  &   352 $\pm$ 7    \\[3pt]
$\bar{\sigma}_g$  [$\rm km~s^{-1}$]$^{n}$ &  560 $\pm$ 15  &   455 $\pm$ 55    \\[3pt]
\hline
\hline
\end{tabular}
\end{center}
\tablefoot{
$^{a}$ Line peak wavelength from the Gaussian modelling of Ly$\alpha$- and CIV-CEN spectrum.
$^{b}$ Redshift of the CEN corresponding to $\rm \lambda_{CEN}$. We used as Ly$\alpha$ and CIV $\rm \lambda$ rest-frame 1215.67 \AA~and 1549.48 \AA, respectively.
$^{c}$ Maximum projected physical size derived from the 3D-mask.
$^{d}$ Spectral width within which the CEN is detected (from 3D-mask).
$^{e}$ Distance from the quasar position to the SB peak of the CEN.
$^{f}$ Distance from the quasar position to the flux weighted centroid of the CEN.
$^{g}$ Asymmetry parameter, i.e. the ratio between the semi-minor and semi-major axis of the SB map.
$^{h}$ Position angle East of North of the major axis of the SB map.
$^{i}$ Total flux derived from the SB maps (Figs.~\ref{figure:SBLya} and \ref{figure:SBCIV}).
$^{l}$ Total CEN luminosity.
$^{m}$ Averaged velocity dispersion measured from the SNR$\geq$3 velocity dispersion maps (see Section~\ref{sec:kinematic}) of the Ly$\alpha$ and CIV. $^{n}$ Velocity dispersions derived from the Gaussian fit of the total Ly$\alpha$- and CIV-CEN spectrum.
$^{\star}$ Errors according to the spatial resolution of MUSE data.
}
\end{table}

\subsection{Ly$\alpha$ nebula} \label{sec:lyanebula}
We found a Ly$\alpha$-CEN consisting of $\sim$55000 connected voxels by using $SNR_{th}=$2.5 and $N^{vox}_{min}=10000$ pixels. This CEN exhibits a maximum angular extension of $\sim$20~arcsec ($\sim$150~kpc). 
The line emission integrated over the entire CEN spans a maximum wavelength range of $\sim95$ \AA\ ($\sim$ 5000 $\rm km~s^{-1}$) and shows a peak measured through a Gaussian fit at $\rm \lambda_{Ly \alpha} ^{CEN} = (5543.8 \pm 1.8)$ \AA . This corresponds to a redshift of $\rm z_{CEN} ^{Ly\alpha}$ = 3.560$ \pm$ 0.002.

Fig.~\ref{figure:SBLya} shows the SB map of the Ly$\alpha$-CEN. The nebular emission is the region enclosed in the black thick contour, i.e. the projected boundary derived from the 3D-mask which encloses the detected CEN with $\rm SNR> 2.5$.
The thin contours indicate the SNR levels derived from the SNR map.
This Ly$\alpha$-CEN exhibits a roughly symmetric shape on large scales suggesting a circular (or mildly elliptical) morphology with marked deviations at small radii. In order to obtain a more quantitative description of its morphology we computed the flux weighted centroid shift ($\rm d^{cen}_{CEN}$), the peak displacement ($\rm d^{peak}_{CEN}$), the asymmetry ($\alpha$) and the position angle East of North ($\phi$) of the CEN. The first two quantities were measured with respect to the quasar position, while the remaining ones are estimated from the Stokes parameters as in AB19 (see Table~\ref{table:NebulaProp}).
We found general consistency with morphological properties reported by AB19 in the QSO MUSEUM sample. Indeed, both the flux-weighted centroid shift and the asymmetry measured for the Ly$\alpha$-CEN around J1538+08 are close to the median values reported in AB19, i.e. 8.3 kpc and 0.71 respectively.
The Ly$\alpha$-CEN exhibits a maximum SB value of  $\rm \sim10^{-16}~erg~s^{-1}~cm^{-2}~arcsec^{-2}$~ and a total luminosity of $\rm L _{Ly \alpha} = (2.06 \pm 0.02) \times 10^{44}$ \lumcgs. 

\begin{figure}[t]
   \begin{center}
   \includegraphics[height=0.25\textheight,angle=0]{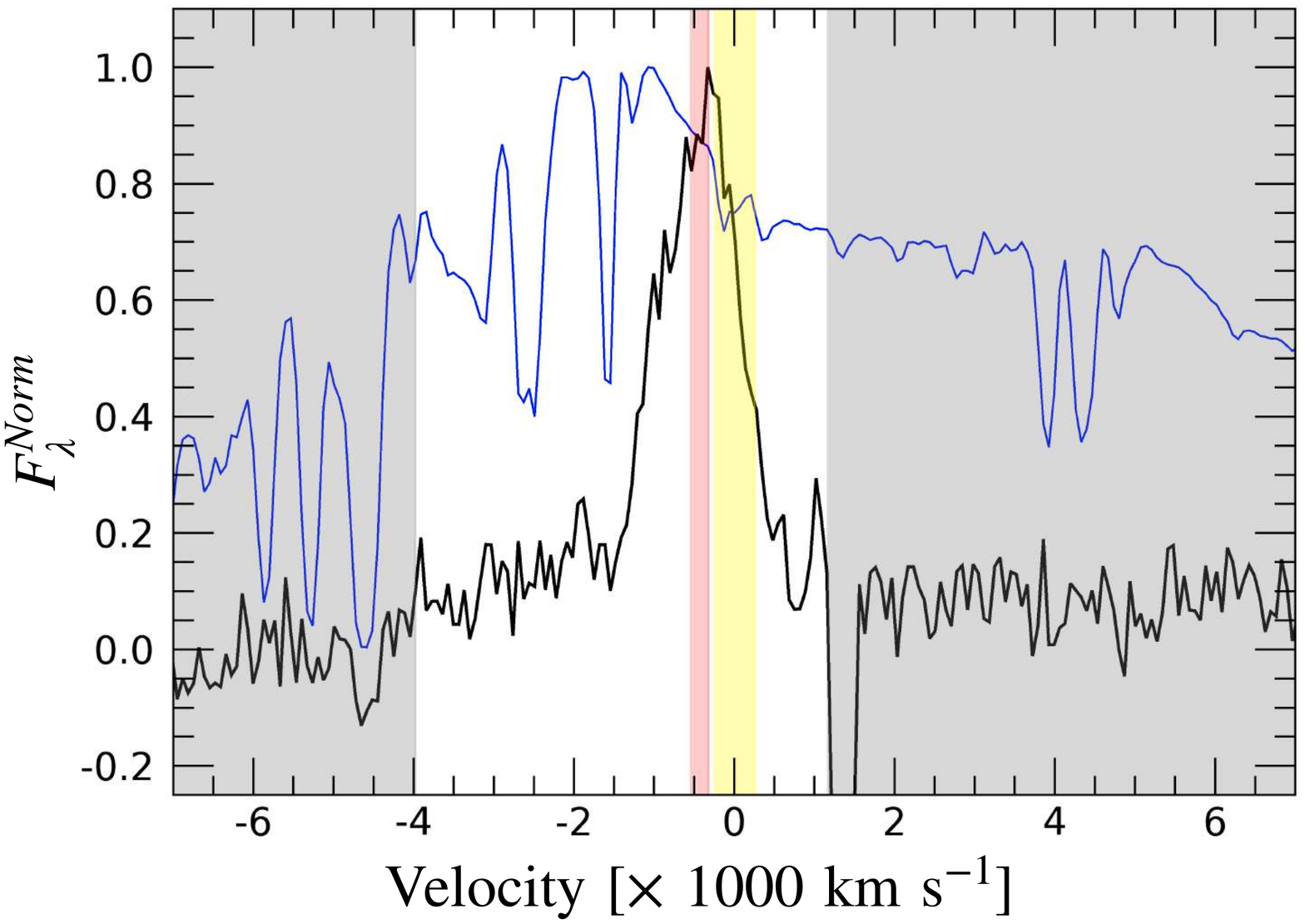}
   \caption{Black and blue represents the Ly$\alpha$ emission lines in the CEN and quasar spectrum, respectively. These Ly$\alpha$ line spectra are reported in velocity space relative to the $\rm z_{QSO}$ and they both are normalized to their peak. The white region reports the spectral range of the 3D-mask. The red area marks the peak wavelength of the Ly$\alpha$ line derived through Gaussian modelling including the errors. The yellow area indicates the 1$\sigma$ uncertainty of the systemic redshift of the quasar \citep{Vietri18}. The absorption line at $\sim$1500 $\rm km~s^{-1}$ is a sky feature which could not be appropriately removed by the algorithm \texttt{CubePSFSub}.}
   \label{figure:Lyal}
   \end{center}
\end{figure}

The comparison between the Ly$\alpha$ spectra of the quasar and the CEN normalized to their peak emission is shown in Fig.~\ref{figure:Lyal}. The quasar spectrum was extracted from an aperture with radius of 3 arcsec, while the CEN spectrum was obtained from the PSF- and continuum- subtracted datacube by summing the spectral contribution of all the spaxels belonging to the CEN (i.e. from the 3D-mask). The x-axis reports the velocity relative to $\rm z_{QSO}$.
The white area in the plot corresponds to the spectral width where the CEN was detected (i.e. the maximum spectral extent of the 3D mask; see also Table~\ref{table:NebulaProp}). The red area marks the peak of the Ly$\alpha$ profile of the CEN and includes the errors. 
The measure of the peak has been obtained through a Gaussian modelling plus a constant continuum term, performed within a range $\pm 500$~\AA\ (or $\pm$ 27,000 $\rm km~s^{-1}$) from the Ly$\alpha$ wavelength at $\rm z_{QSO}$. This peak appears to be reasonably (i.e. at $\sim$1$\sigma$) consistent with the $\rm z_{\rm QSO}$ and its 1$\sigma$ uncertainty (yellow area in Fig.~\ref{figure:Lyal}).

\begin{figure}[t]
   \begin{center}
   \includegraphics[width=0.48\textwidth]{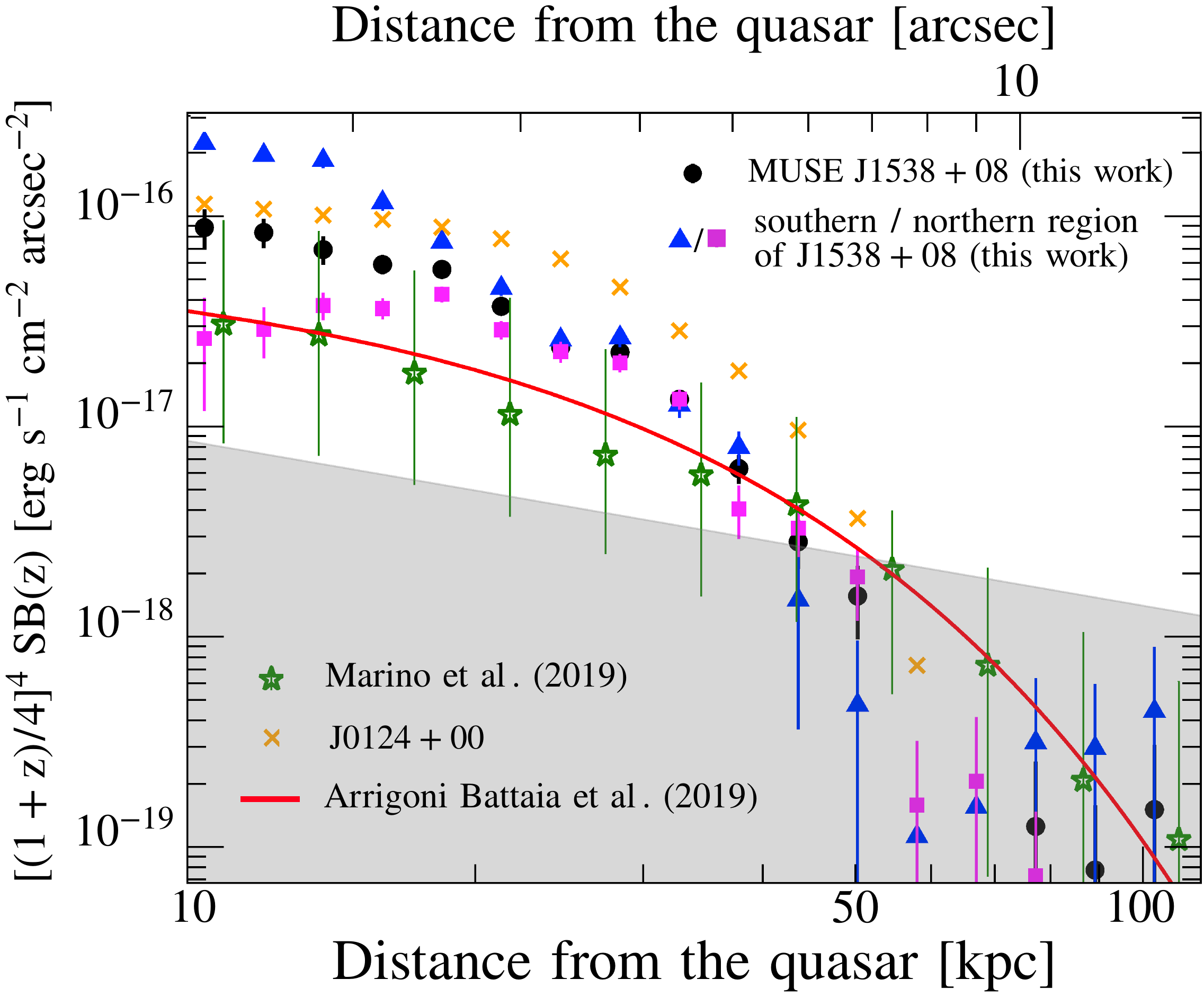}
   \caption{Ly$\alpha$ SB radial profile (black circles) of the pseudo-NB image. The points represent bins of SB measured in concentric annuli centered on the quasar position and their errors indicate the 16th and 84th percentile (corresponding to $\sim$68$\%$ of the SB distribution centered on the median value, i.e. $\pm$ 1$\sigma$ for a Gaussian distribution). The bin width is chosen to be uniformly spaced in a logarithmic scale. The blue triangles and magenta squares show the radial profiles measured in the southern and northern region of the CEN, respectively (see Fig.~\ref{figure:SBLya}). The red curve is the modelled average radial profile of the AB19 quasar sample. We also include the average radial profile of the Ly$\alpha$-CEN of the sample of B16 recalculated by \cite{Marino19} (the error bars represent the 25th and 75th percentiles).We report as orange crosses the radial profile of the Ly$\alpha$-CEN around the WISSH quasar J0124+00 in the B16 sample. All the SB profiles are corrected for the cosmological dimming effect and re-scaled at z=3 (i.e. $\rm [(1+z)/4]^4 \times SB (z)~$). The grey shaded area indicates the 2$\sigma$ Poisson noise relative to the rescaled profiles of the Ly$\alpha$-CEN around J1538+08 (see Section \ref{sec:lyanebula} for details). We verified that the decline of the SB radial profile from the northern region (magenta squares) at short radius towards the center is not due to the white region immediately north to the quasar (see caption Fig.~\ref{figure:SBLya})}
   \label{figures:rpLya}
   \end{center}
\end{figure}

Fig.~\ref{figures:rpLya} reports the circularly averaged radial SB profile of the Ly$\alpha$-CEN in J1538+08 (black points) corrected for minimal residual background emission\footnote{Similarly to B16, the SB is estimated at scales much larger than the CEN and the profile is corrected for possible residual background due to unresolved/faint background or foreground (see Appendix \ref{Saturstar}) emission not removed during the continuum subtraction. This subtraction, performed for both the Ly$\alpha$ and the CIV profiles (see Section~\ref{sec:CIV}), was measured to be negligible accounting for 5-15 $\%$ of the flux of the most external annuli (used for the SB profile) placed at the boundary of each CEN.}. The profile was extracted from concentric annular regions centered on the quasar position in the image obtained by collapsing the spectral region in the CEN 3D-mask.
The grey area indicates the $2 \sigma$ Poisson noise\footnote{The Poisson noise was estimated as the average value of the standard deviations within background regions $1" \times 1"$ of the pseudo-NB divided by the square root of each annulus area.}. 
The radial profile was compared with the average radial profiles reported by B16 and AB19. All the SB radial profiles in Fig.~\ref{figures:rpLya} were corrected for the cosmological $\rm (1+z)^4$ SB dimming factor and re-scaled at z=3 for comparison.

We also report the rescaled profile of the CEN around the quasar J0124+00 (which also belongs to the WISSH quasar sample, e.g. \citealt{WISSHII}), which was identified by B16 as the brightest and most peculiar one in their sample in terms of sharp-peaked morphology, large velocity dispersion and SB radial profile.
Both the Ly$\alpha$-CEN around these WISSH quasars exhibit a central (i.e. <40-50~kpc) SB excess relative to the B16 and AB19 average profiles.

We further investigated the origin of this SB excess in J1538+08 by comparing the radial profiles extracted for two regions, i.e. south and north of the quasar, which are delimited by the green dashed lines in Fig.~\ref{figure:SBLya}. Accordingly, the southern region includes the SB peak of the Ly$\alpha$-CEN  and the CIV-CEN (see Fig.~\ref{figure:SBCIV}). The SB excess is clearly due to the contribution from the southern region (blue triangles in Fig.~\ref{figures:rpLya}) at $\leq$20 kpc, while both regions equally contribute to the SB excess at larger radii.
%

\subsection{CIV nebula} \label{sec:CIV}
We scanned the datacube searching for diffuse emission in additional ionic transitions typically probing the CGM. The only extended emission detected at good significance level is traced by the CIV line. Given its typically lower emissivity compared with the Ly$\alpha$ we started our search by setting $SNR_{th}=$2.5 and $N^{vox}_{th}=$10000 and then progressively lowered $N^{vox}_{th}$ by steps of 1000 voxels until the extended emission was detected. We found that the $N^{vox}_{th}=4000$ was required to detect this CEN, that indeed consists of 4726 voxels. 

\begin{figure}[!t]
   \begin{center}
   \includegraphics[height=0.38\textheight,angle=0]{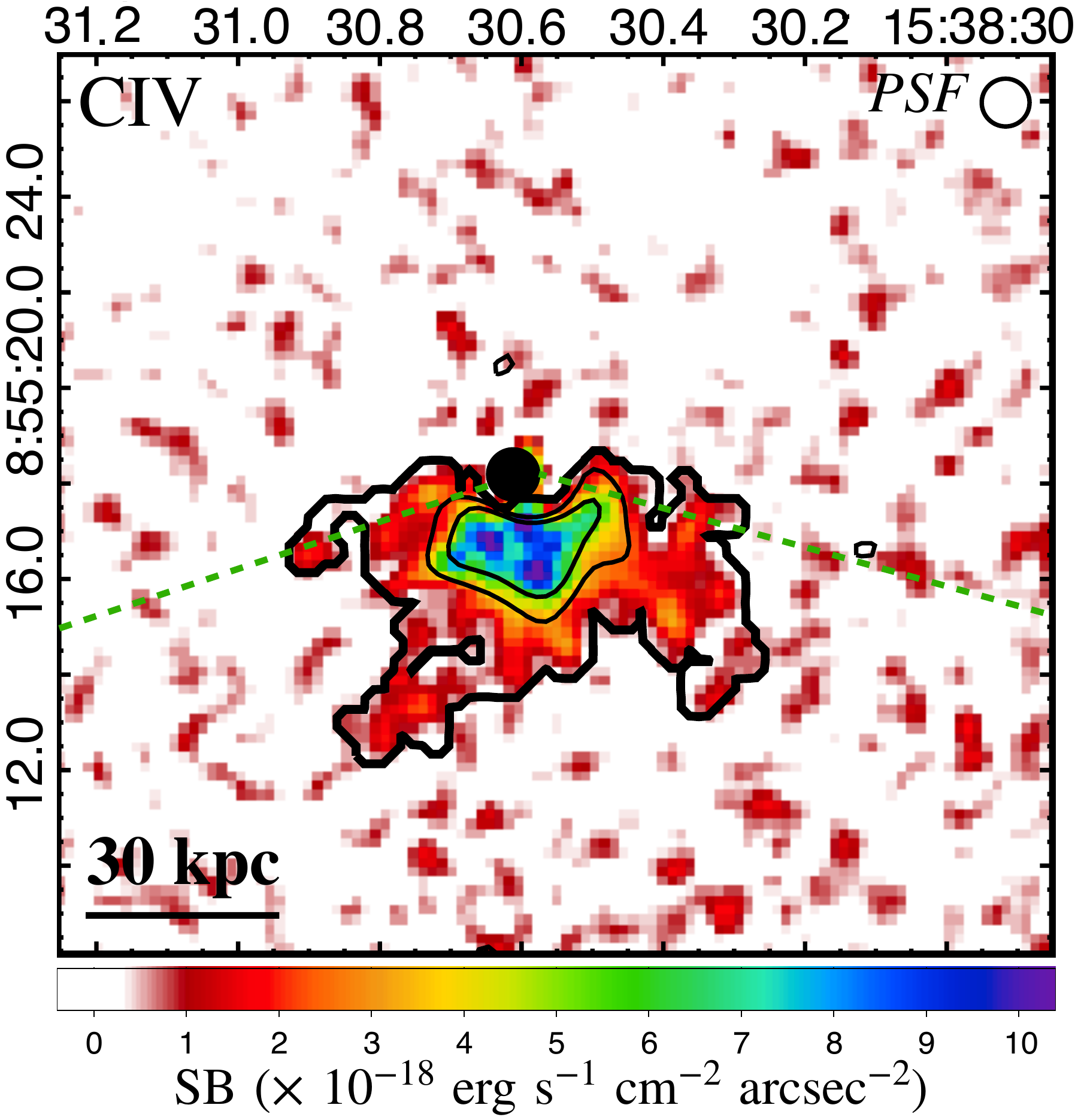}
   \caption{Optimally extracted SB map of the CIV-CEN. The black dot, the contours and the dashed green lines have the same meaning of Fig.~\ref{figure:SBLya}. Thin contours reports levels of $\rm SNR$=3 and 5.}
   \label{figure:SBCIV}
   \end{center}
\end{figure}

The SB map of the CIV-CEN is shown in Fig.~\ref{figure:SBCIV}. The CIV-CEN is much smaller than the Ly$\alpha$-CEN, with a maximum angular extension of $\sim 10$~arcsec, which corresponds to a projected physical size of $\sim$75~kpc. It is almost completely contained in the southern region and exhibits an asymmetric morphology with $\alpha=0.62$. The peak of the $\rm SB_{CIV}$ is located roughly at the same position of the $\rm SB_{Ly \alpha}$ peak, at a distance of $\sim 11.0 \pm 3.5~\rm kpc$ from the quasar position. The value of the SB peak ($\rm \sim10^{-17}~erg~s^{-1}~cm^{-2}~arcsec^{-2}$) and the luminosity of the CIV-CEN ($\rm L _{CIV} = (9.33 \pm 0.33) \times 10^{42} ~\rm{erg~s^{-1}}$) are one order of magnitude lower than those measured for the Ly$\alpha$-CEN.

\begin{figure}
   \begin{center}
   \includegraphics[height=0.25\textheight,angle=0]{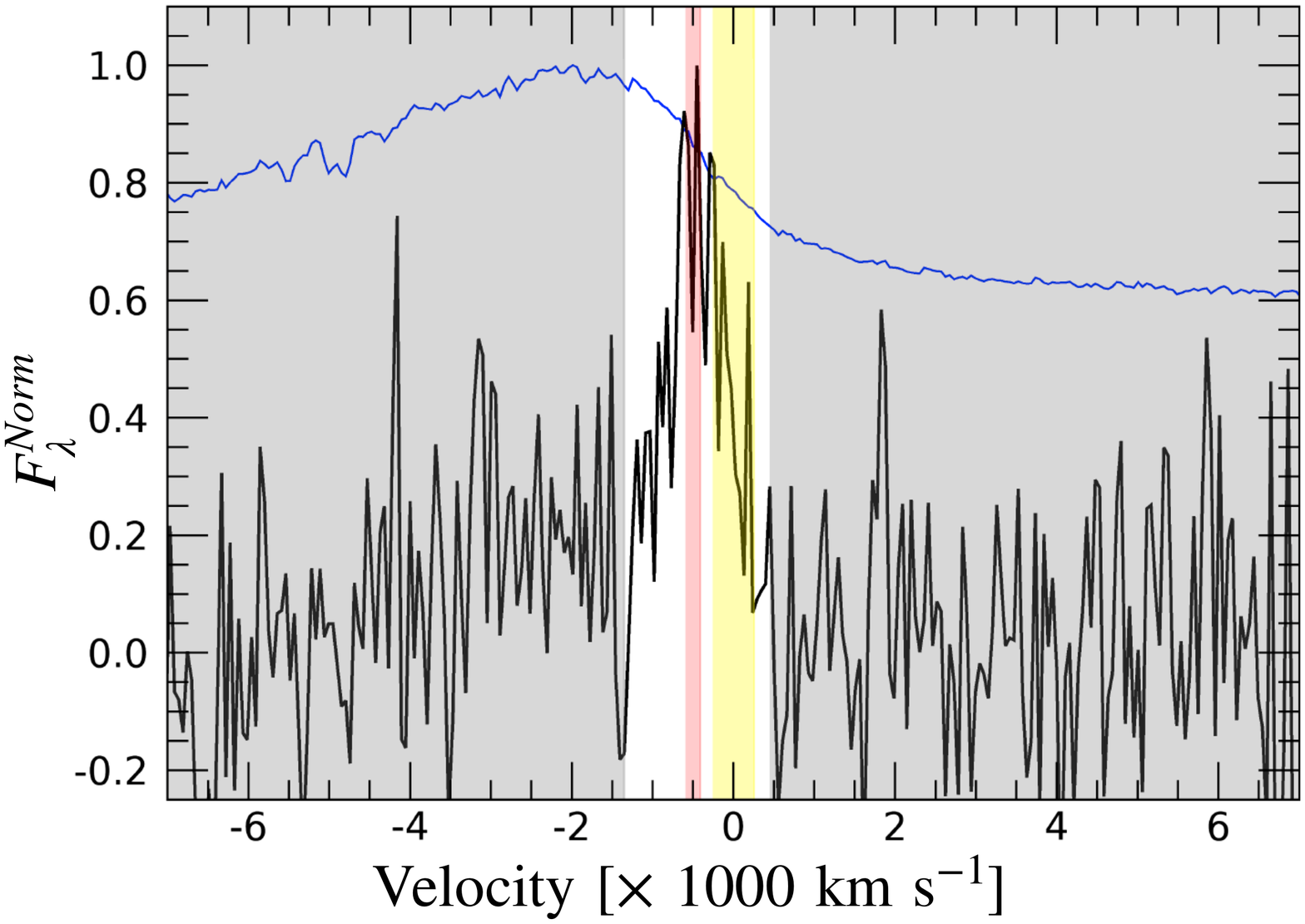}
   \caption{Spectrum of the CIV emission as a function of the velocity with respect to the quasar redshift of the CIV-CEN (black) and quasar (blue) normalized to their peak emission. Labeling and symbols as in Fig.~\ref{figure:Lyal}}
   \label{figure:CIVl}
   \end{center}
\end{figure}

\begin{figure}
   \begin{center}
   \includegraphics[width=0.46\textwidth]{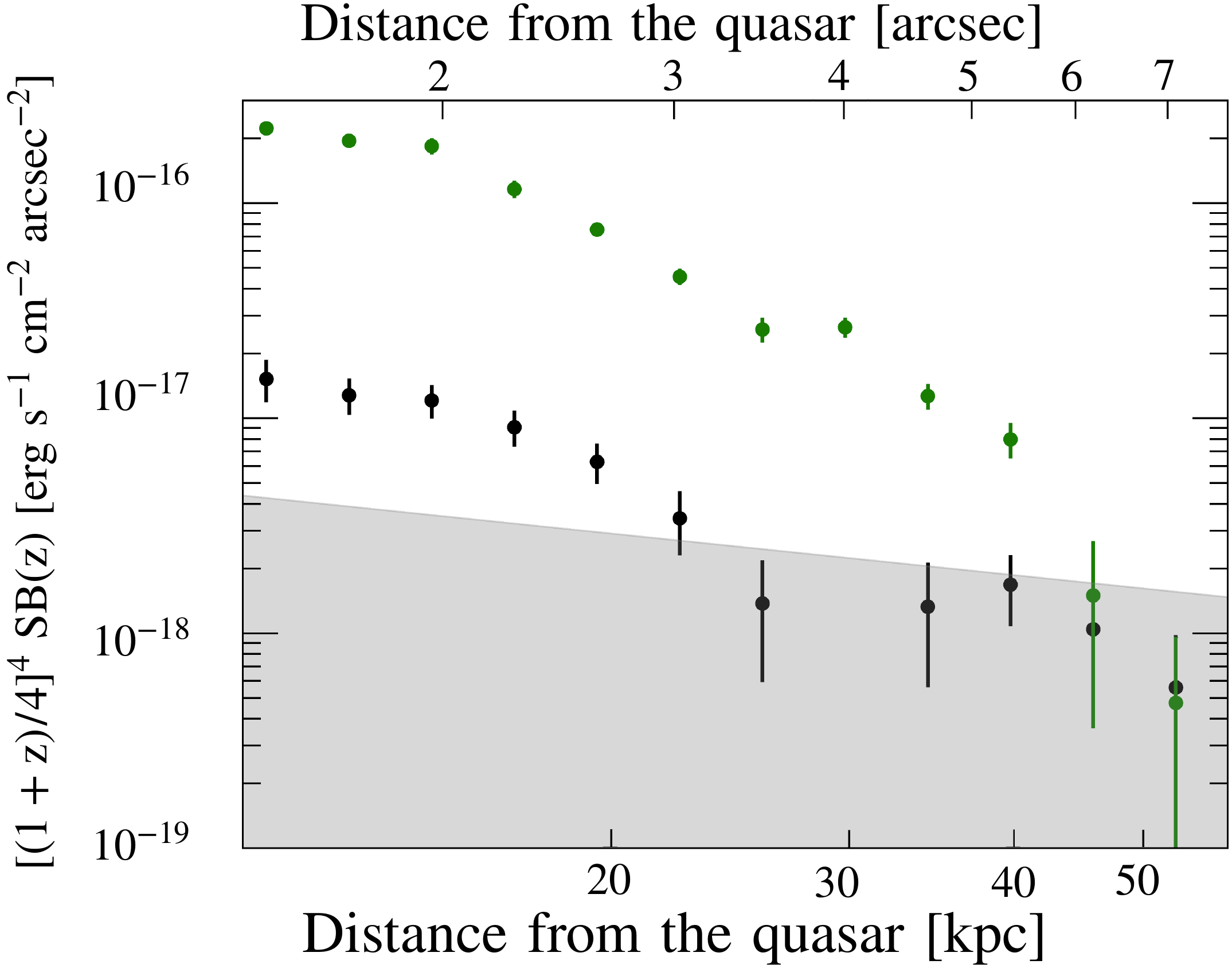}
   \caption{SB radial profile of the CIV-CEN (black) extracted in the southern region (see Fig.~\ref{figure:SBCIV}). The SB profile for the Ly$\alpha$-CEN, extracted in the same region, is also reported in green. These profiles are re-scaled at z=3 and corrected for the cosmological dimming effect. For other details refer to the caption of Fig.~\ref{figures:rpLya}.}
      \label{figure:radproCIV}
\end{center}
\end{figure}

Fig.~\ref{figure:CIVl} shows the comparison between the CIV emission line in the CEN and the quasar spectrum.
The CIV-CEN was detected in a relatively narrow spectral range ($\sim 2000~\rm{km~s^{-1}}$) compared to the spectral region of the Ly$\alpha$-CEN ($\sim 5000~\rm{km~s^{-1}}$). 
Through a Gaussian modelling of the line (similarly to the one performed for the Ly$\alpha$), we found that the CIV line peaks at $\rm \lambda _{CIV} ^{CEN} = 7064.7 \pm 1.8$~\AA. This corresponds to a redshift of $\rm z_{CEN} ^{CIV}$=3.559 $\pm$ 0.001, i.e. in good agreement with the Ly$\alpha$-CEN one. We computed the SB radial profile of the CIV emission (Fig.~\ref{figure:radproCIV}) in the southern region.  
The emission of the CEN is  clearly above the Poisson noise up to $\sim$ 20 kpc from the quasar position.
\begin{figure*}
   \begin{center}
   \includegraphics[width=0.81\textwidth]{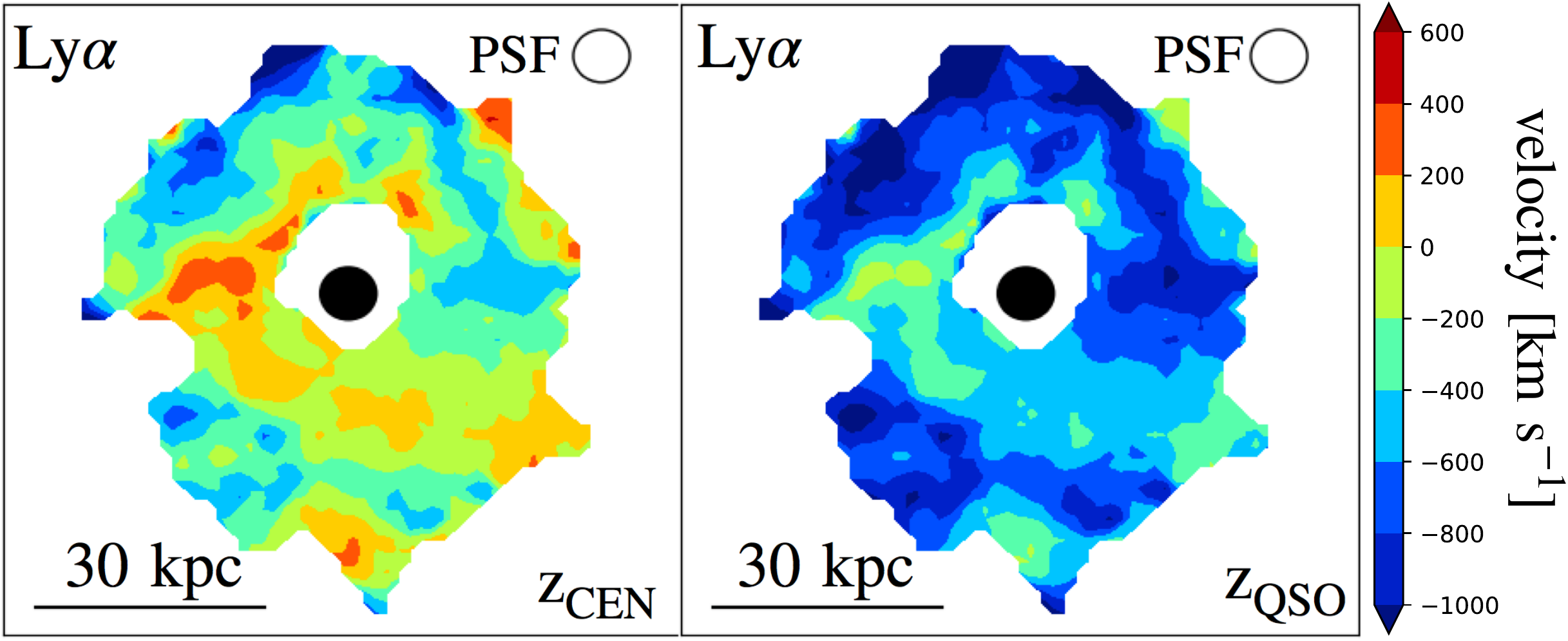}
   \caption{Flux-weighted velocity maps of the Ly$\alpha$-CEN for SNR$\geq$3 regions relative to $\rm z_{CEN}$ (left) and to $\rm z_{QSO}$ (right).}
   \label{image:Lyavelocity}
   \end{center}
\end{figure*}
\begin{figure*}
   \begin{center}
   \includegraphics[width=0.8\textwidth]{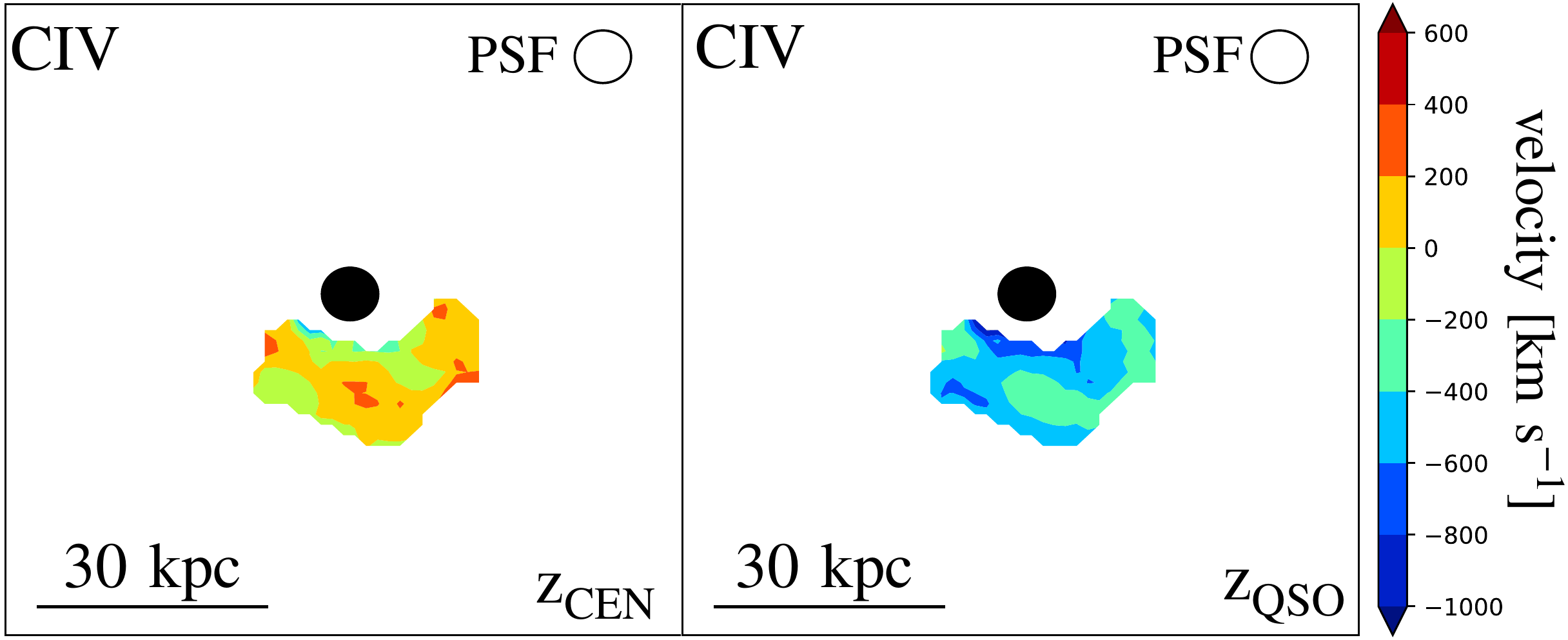}
   \caption{Flux-weighted velocity maps for SNR$\geq$3 regions of the CIV-CEN. As in Fig.~\ref{image:Lyavelocity} the right and the left panels show the velocity relative to $\rm z_{CEN}$ and to $\rm z_{QSO}$, respectively.}
   \label{image:CIVvelocity}
   \end{center}
\end{figure*}

\subsection{Kinematic properties of the CEN} \label{sec:kinematic}

\begin{figure}
   \begin{center}
   \includegraphics[width=0.48\textwidth]{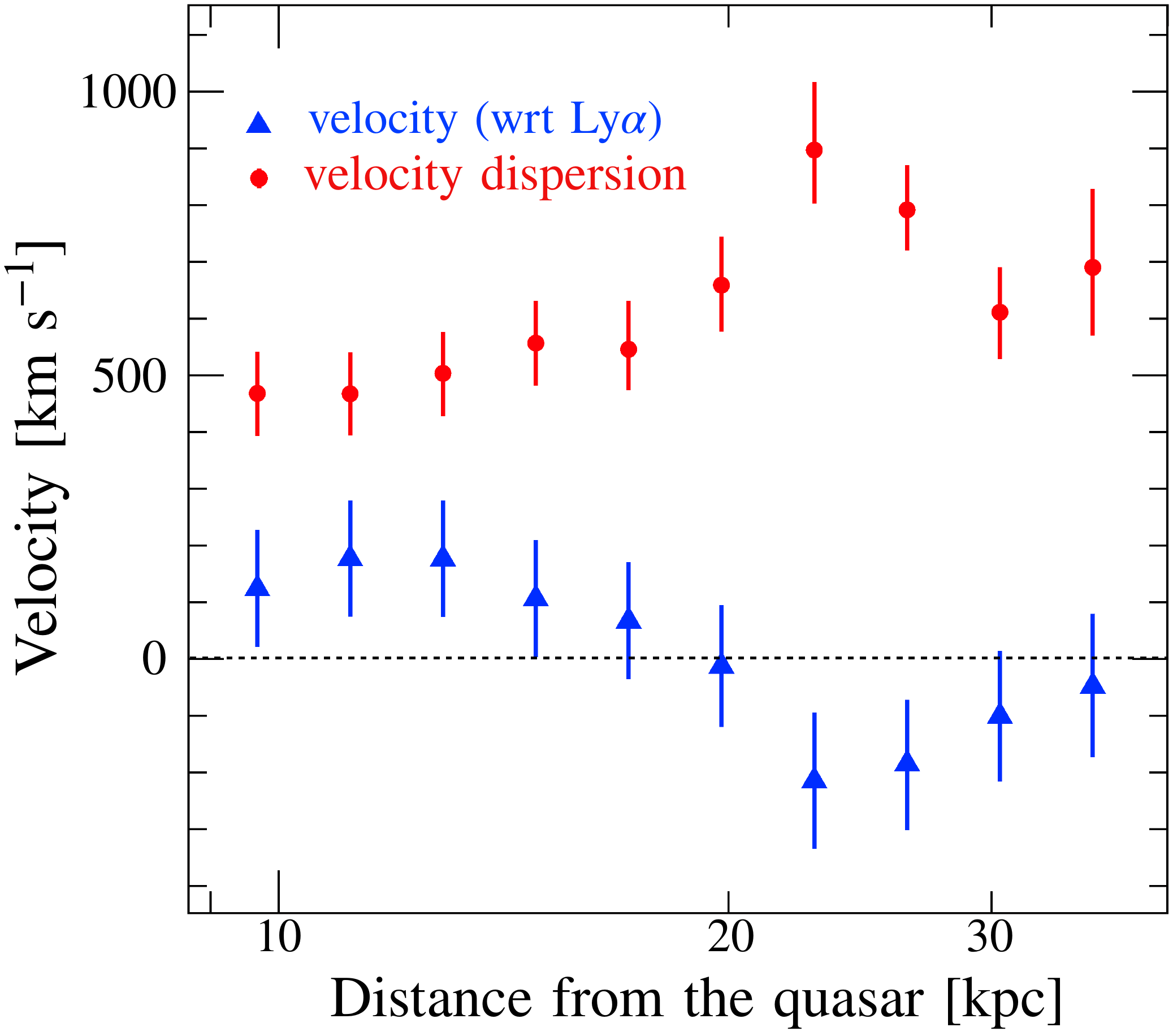}
   \caption{Radial profile of velocity (blue triangles) and dispersion (red circles) derived from the modelling of a single Gaussian of the Ly$\alpha$ profiles extracted in concentric annular region centered on the quasar position.}
   \label{image:RPdispvel}
   \end{center}
\end{figure}
\begin{figure*}[ht]
   \begin{center}
   \includegraphics[width=0.84\textwidth]{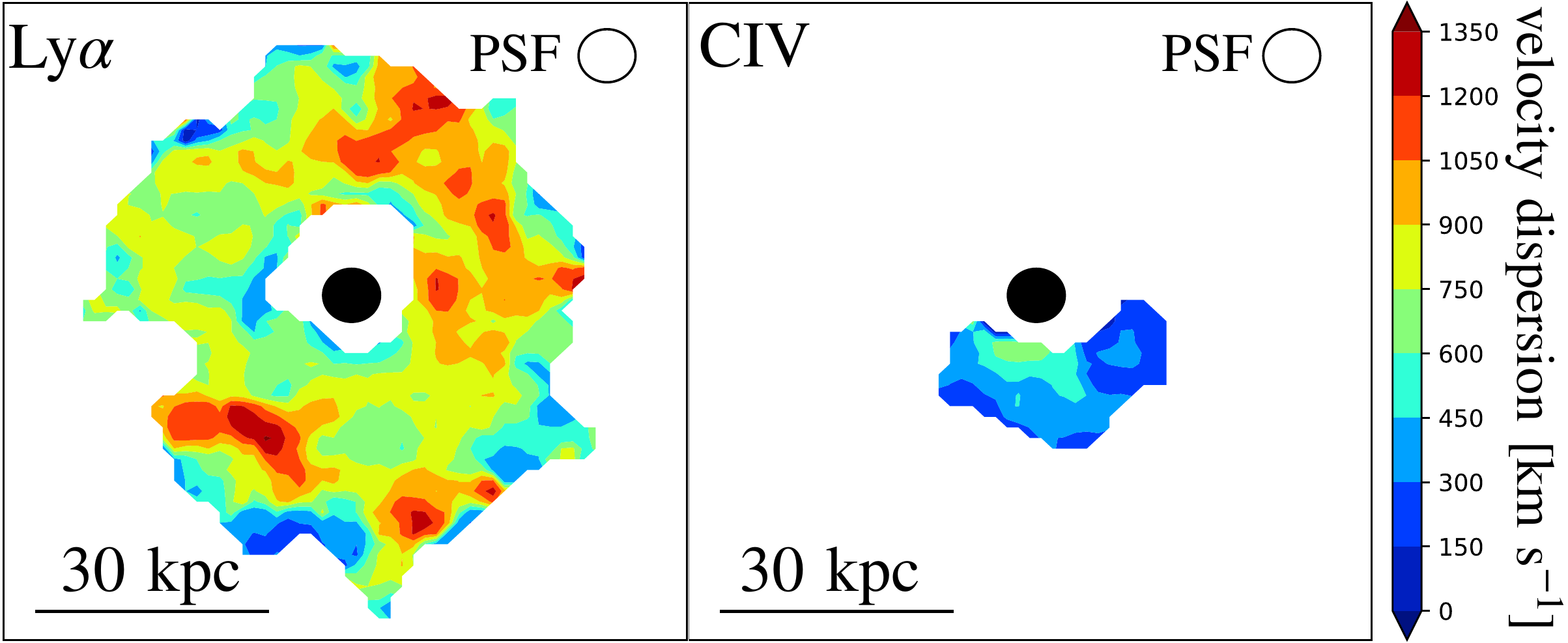}
   \caption{Flux-weighted velocity dispersion maps for Ly$\alpha$- (left) and CIV- (right) CEN. The MUSE spectral resolution in these dispersion maps is $\approx$70~$\rm km~s^{-1}$.}
   \label{image:dispersions}
   \end{center}
\end{figure*}

Figs.~\ref{image:Lyavelocity} and \ref{image:CIVvelocity} show the flux-weighted velocity maps in regions with SNR$\geq$3 for the Ly$\alpha$- and CIV-CEN, respectively. Each panel reports the velocities relative to the CEN ($\rm z_{CEN}$; left panel) and $z_{\rm QSO}$ (right panel) rest-frame, respectively. There is no significant structure indicative of bulk motion or clear separation between red/blueshifted gas components. 
Interestingly, almost the entire Ly$\alpha$-CEN exhibits velocities spanning $\sim 1000 ~\rm km~s^{-1}$. 

We also investigated the radial velocity profile of the Ly$\alpha$-CEN to assess a possible trend with the distance from the quasar. Specifically, we extracted spectra from the PSF- and continuum- subtracted datacube in each annular region used for the SB radial profiles (see Fig.~\ref{figures:rpLya}). For each spectrum, we modelled the Ly$\alpha$ emission line with a Gaussian profile. 
Fig.~\ref{image:RPdispvel} shows the radial distribution of the best-fit velocities relative to the $\rm z_{CEN}$ (blue points). The velocity profile exhibits a constant value around $\sim100-200~\rm km~s^{-1}$ up to 15-20~kpc. There is an apparent decline of the velocity down to $-200~\rm km ~s^{-1}$ at larger radii (20-30~kpc).

The flux-weighted velocity dispersion maps for Ly$\alpha$- (left panel) and CIV-CEN (right panel) are shown in Fig.~\ref{image:dispersions}. For the Ly$\alpha$-CEN we do not notice any specific gradient as the map shows a nearly constant $\rm \sigma _v$ in the range 600-900 $\rm km~s^{-1}$ with few regions reaching values larger than 1000$~\rm km~s^{-1}$. In case of the CIV-CEN the map shows a factor of 2 lower velocity dispersions than the Ly$\alpha$-CEN.
Tab.~\ref{table:NebulaProp} lists the average velocity dispersions ($\bar{\sigma}_v$) obtained from these maps. The dispersion estimated through Gaussian model fitting ($\bar{\sigma}_g$) of the total spectrum of the Ly$\alpha$- and CIV-CEN is also reported.

In order to better explore the possible presence of a radial gradient of the \lya emission velocity dispersion, we computed the radial dispersion profile using the same methodology adopted for the velocity profile (see red points in Fig.~\ref{image:RPdispvel}).
The profile exhibits $\sigma _{\rm v} \sim 500~\rm km~s^{-1}$ up to $\rm \sim 15-20~kpc$ and and increase up to $\sigma _{\rm v} \sim900~\rm km ~s^{-1}$ at $\rm \sim 25 ~kpc$ from the quasar. The profiles could only be accurately computed up to 40 kpc since the line profile modelling at larger distances is completely unconstrained.

\subsection{CIV/\lya and HeII/\lya line ratios} \label{sec:ratioCIVLya}

We derived a CIV/\lya line flux ratio of $0.06 \pm 0.01$, in agreement with the approximate values previously estimated for some tentative ($\sim 2.2 \sigma - 2.8 \sigma$) detections of CIV-CEN in B16. Similarly to our result, the B16 values were obtained by estimating the total flux of the voxels associated with the Ly$\alpha$ 3D-Mask shifted at the wavelength of the CIV corresponding to $\rm z_{CEN}$, divided by the total flux of the Ly$\alpha$-CEN (reported in Table~\ref{table:NebulaProp}). This procedure avoids the bias due to aperture effects (i.e. different extensions for different line emissions) and provides a conservative estimate if, as expected, the Ly$\alpha$-CEN is the brightest and more extended one.

\begin{figure}[!t]
   \begin{center}
   \includegraphics[width=0.49\textwidth]{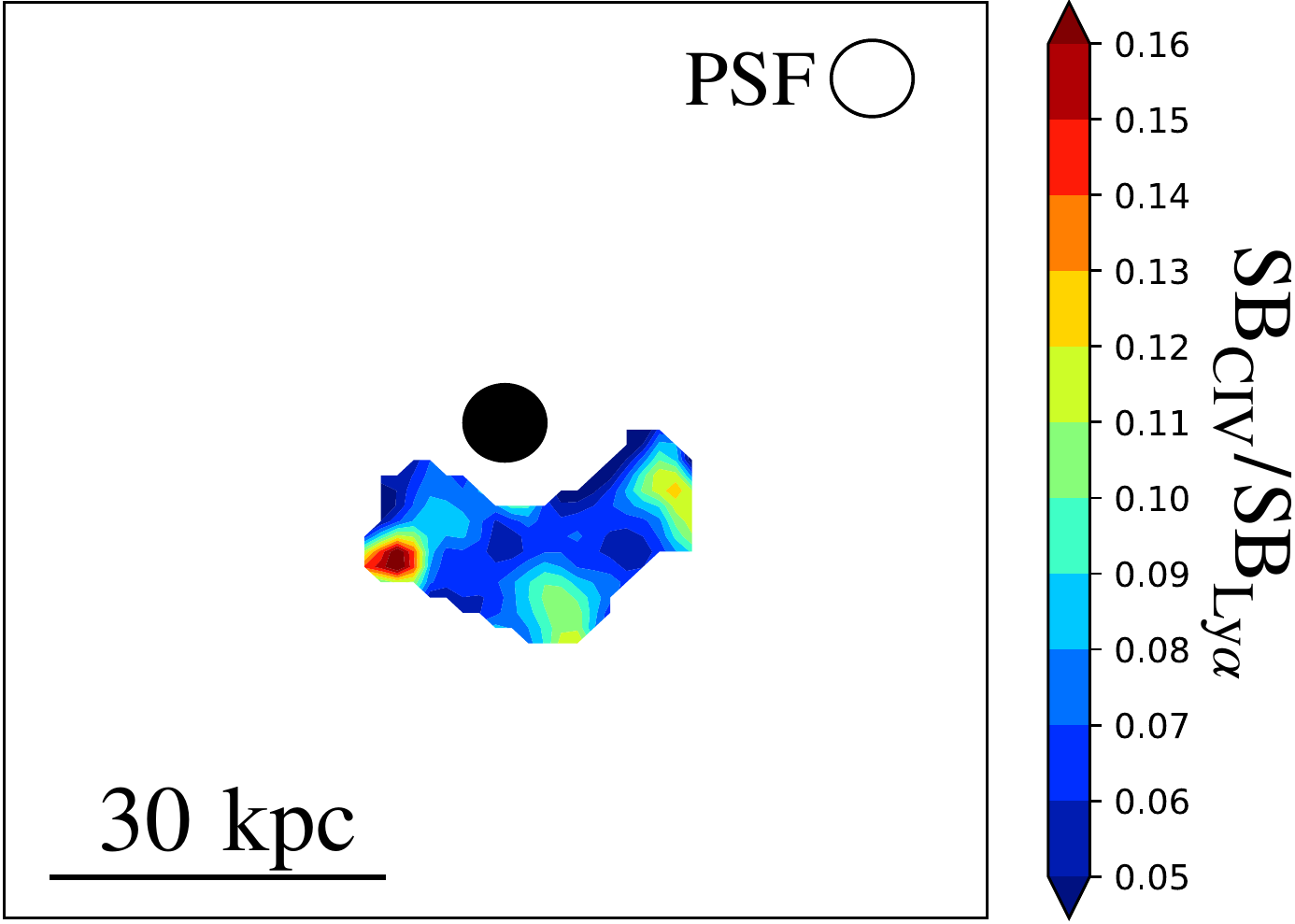}
   \caption{Map of the CIV/Ly$\alpha$ SB ratio. Only regions with SNR$>3$, for both the Ly$\alpha$- and CIV-CEN, are displayed.}
   \label{fig:ratioLyaCIV}
   \end{center}   
\end{figure}

The CIV/Ly$\alpha$ ratio map reported in Fig.~\ref{fig:ratioLyaCIV} is the ratio between the CIV and Ly$\alpha$ SB maps over the SNR$\geq$3 regions. The spatial distribution of the ratios suggests that the metal distribution is not completely homogeneous.
The median value of the CIV/Ly$\alpha$ ratio directly measured from the CIV/Ly$\alpha$ map is $0.08^{+0.11}_{-0.04}$, where the uncertainties are derived by the 16th and 84th percentiles (corresponding to $\pm$1$\sigma$ for a Gaussian distribution).
A consistent value was obtained by computing the SNR-weighted average of the map values of both Ly$\alpha$- and CIV-CEN.

No HeII emission line is visible in the quasar spectrum and no HeII-CEN is detected around J1538+08 at $\rm z_{CEN}$.
This is not surprising: while extended HeII emission is commonly observed around high-redshift radio galaxies (e.g. \citealt{Wilman00}, \citealt{Villar07}), it is only found in 6$\%$ of the RQQs in the B16 sample.  We inferred a 2$\sigma$ upper limit to the HeII/Ly$\alpha$ ratio of $<0.02$ by applying the same Ly$\alpha$ 3D-mask shift procedure used to obtain the total CIV/Ly$\alpha$ flux ratio.

\subsection{Asymmetry in Ly$\alpha$ line profile} \label{sec:skew}

\begin{figure}[t]
   \begin{center}
   \includegraphics[width=0.5\textwidth]{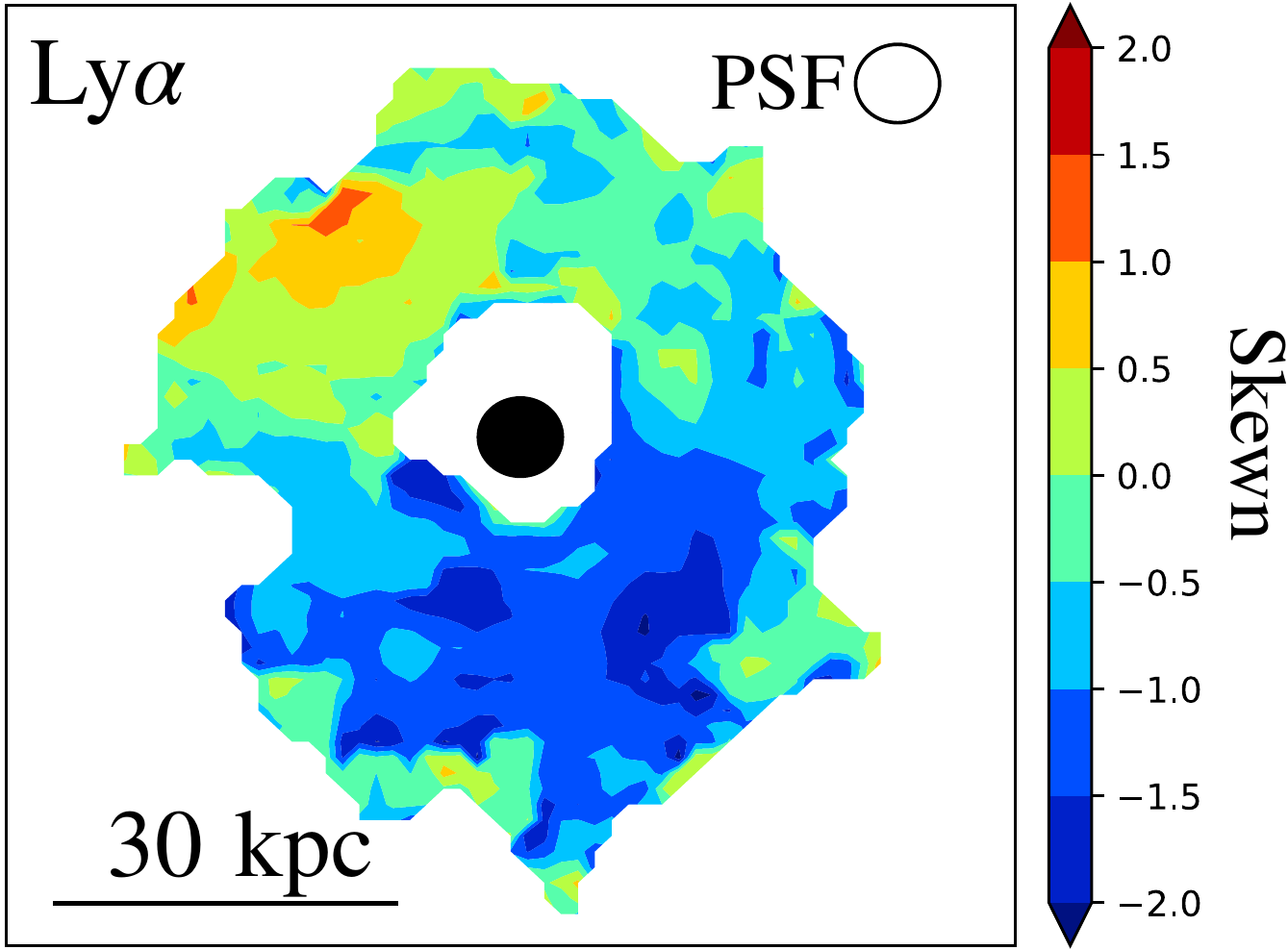}
   \caption{Skewness map of the line profile of the Ly$\alpha$-CEN.}
   \label{image:skew}
   \end{center}
\end{figure}

\begin{figure}[ht]
   \begin{center}
   \includegraphics[width=0.48\textwidth]{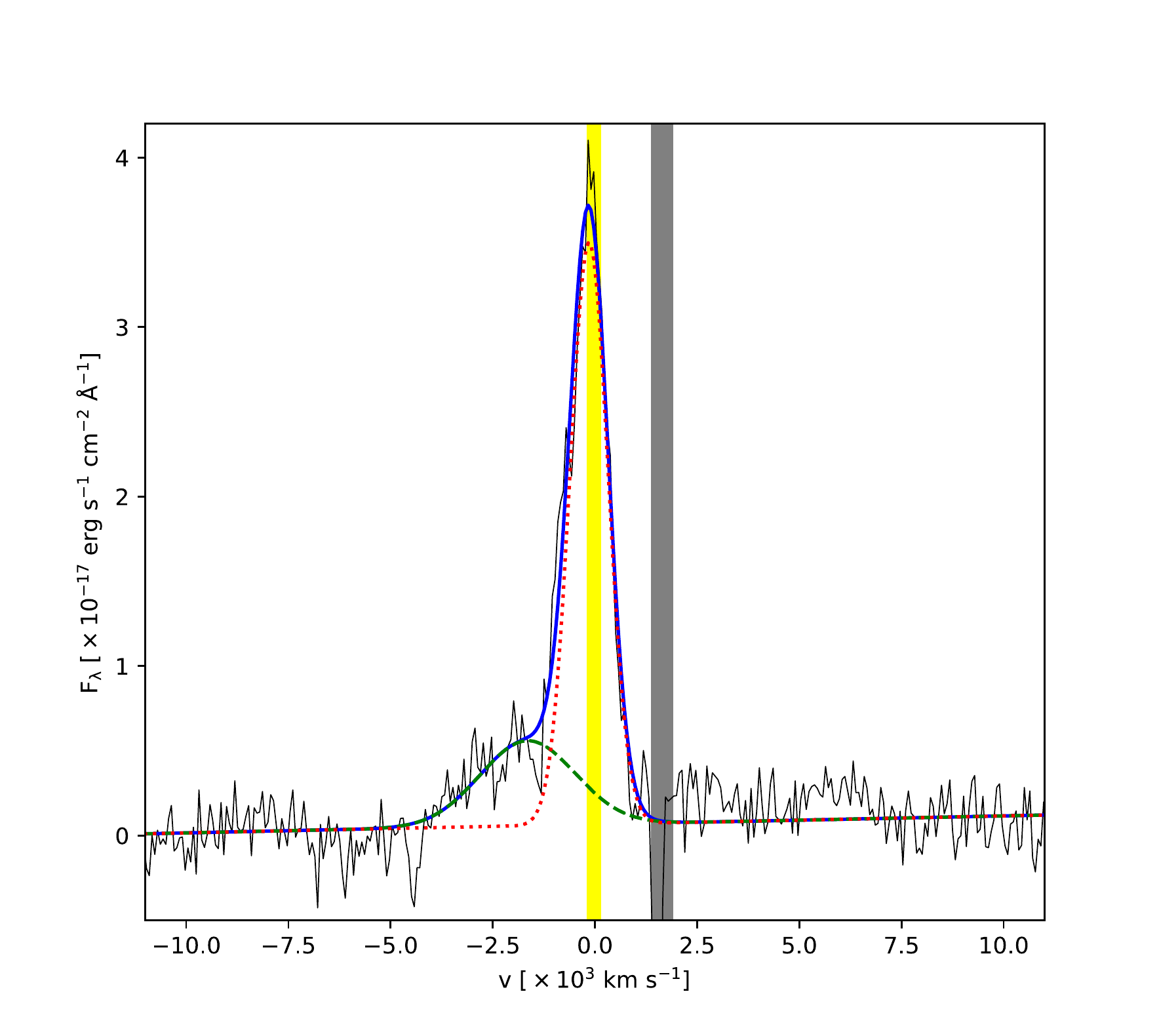}
   \caption{Ly$\alpha$ spectrum extracted from the CEN region with $sk \rm <-0.5$ and $\rm SNR >3$ and modelled (blue solid line) by two Gaussian components (red dotted and green dashed lines). Interestingly, the broad ($\rm \sim 1170~km~s^{-1}$) wing is blueshifted by $\rm \sim 1520~km~s^{-1}$ from the narrow component. The grey area is not considered in the modelling as it includes a region contaminated by a sky feature. The yellow region reports the $\pm1\sigma$ uncertainty on the $z_{\rm QSO}$.}
   \label{image:FitLya}
   \end{center}
\end{figure}

The integrated profile of the Ly$\alpha$ emission line extracted from the entire CEN and reported in Fig.~\ref{figure:Lyal} exhibits a blue tail which can be modelled by an additional component with negative velocity. 
With the aim of exploring the spatial distribution of this blue tail component, we mapped the asymmetry of the \lya emission line profile by using the skewness estimator ($sk$). 
As for the velocity and dispersion maps, we derived the skewness map as the third moment of the flux distribution according to the following formula:\\
\begin{equation}
    \rm skewness=\frac{\sum _{i,j,k} ^{3D-Mask} \rm{Residuals_{i,j,k} ^{3}} \times F_{i,j,k}}{M_2 ^{3/2} \times \sum _{i,j,k} ^{3D-Mask} F_{i,j,k}}
\end{equation}
where $\sum _{i,j,k} ^{3D-Mask}$ is the sum on each voxel defined in the 3D-Mask, $\rm F_{i,j,k}$ is the PSF and continuum-subtracted flux from the datacube and $\rm M_2$ is the second moment of the flux distribution, defined as: \\
\begin{equation}
    \rm M_2 = \frac{\sum _{i,j,k} ^{3D-Mask} \rm{Residuals_{i,j,k} ^{2}} \times F_{i,j,k}}{[\sum _{i,j,k} ^{3D-Mask} F_{i,j,k} ]^2 - [\sum _{i,j,k} ^{3D-Mask} F_{i,j,k} ^2 ]}
\end{equation}
We found that the southern part of the CEN with skewness values $sk<-0.5$ (blue regions in Fig.\ref{image:skew}) overlaps with the region of high SB. We verified that other regions do not exhibit skewness at significant levels (i.e. mostly symmetric and low SNR line profiles).
The Ly$\alpha$ spectrum extracted from spaxels with $ sk\rm<-0.5$ and $\rm SNR \geq 3$ resulted to be clearly asymmetric (Fig.~\ref{image:FitLya}).
We fit it with two Gaussian components and an additional constant term to account for residual background. 
The two Gaussian components exhibit significantly different width, with the broad ($\rm \sigma_{v}\simeq 1170 \pm $ 260 $\rm~km~s^{-1}$) component blueshifted by $\rm v_{ shift} = 1520~\pm~360~km~s^{-1}$ and the narrow one ($\rm \sigma_{v} \simeq 470 \pm $ 70 $\rm~km~s^{-1}$, see Table~\ref{tab:fit}) tracing the systemic $z_{\rm QSO}$. The velocity dispersion found for the narrow component is comparable to the ones reported in Ly$\alpha$-CEN around high-z RQQs (e.g. B16, AB19). The centroid of the narrow component corresponds to a redshift of $z\sim \rm 3.562 \pm 0.001$, which is consistent with $\rm z_{QSO}$.
\begin{table}[t!]
  \begin{center}
    \caption{Properties derived from the two-component Gaussian fit of the Ly$\alpha$-CEN spectrum extracted from regions with SNR$\geq$3 and $sk<$-0.5.}
    \label{tab:fit}
    \begin{tabular}{c c c}
      \hline\hline 
       Components & narrow & broad  \\[3pt]
      \hline
      $\rm \lambda _{cen}$ [\AA]  & $5546.3 \pm 1.3$  & $5518.3 \pm 6.5$ \\[3pt]
      $\sigma _{\rm v} ~[km~s^{-1}]$ & $470 \pm 70$ & $1170 \pm 260$ \\[3pt]
      $\rm Flux ^{a}$ [$\rm \times 10^{-16}~ erg~s^{-1}~cm^{-2}$] & $7.4 \pm 0.7$ & $2.7 \pm 0.9$ \\[3pt]
      \hline
      $\rm v _{shift} ^{b} $ & \multicolumn{2}{c}{($\rm 1520 \pm 360$) $\rm km~s^{-1}$} \\[3pt]
      $\rm v_{max} ^{c}$  & \multicolumn{2}{c}{$\rm (3860 \pm 870) ~km~s^{-1}$} \\[3pt]
      $ \chi ^2/\rm dof $ & \multicolumn{2}{c}{402/366} \\[3pt]
      \hline
    \end{tabular}
  \end{center}
  \tablefoot{
$^{a}${Integrated fluxes of the narrow and broad Gaussian components.}
$^{b}${Velocity offset between the positions of the broad and narrow components;
$^{c}${Maximum velocity of the blueshifted, broad component defined as $\rm v_{max} = v_{shift} + 2 \sigma _{v}$, where $\rm \sigma _{v}$ is the dispersion derived for the broad component.}

}}
\end{table}
%
%


\section{Discussion} \label{sec:discussion}

\subsection{Comparison with the properties of other Ly$\alpha$-CEN samples} \label{sec:discusSB}

We have reported the discovery of a Ly$\alpha$-CEN around the hyper-luminous RQQ J1538+08, which exhibits a projected size of $\sim$150~kpc and a luminosity of $\rm L _{Ly \alpha} \sim 2 \times 10^{44}~erg~s^{-1}$. Previous MUSE studies at similar redshifts (i.e. $z\sim3-4$) and exposures (0.75-1 hours) reported Ly$\alpha$-CEN around similarly luminous RQQs (B16) and quasars with slightly lower $\rm L_{bol}$ (AB19).   
The Ly$\alpha$-CEN around J1538+08 exhibits a maximum projected size which is similar to the average value ($\sim150$~kpc) found for the B16 sample.
The $\rm SB_{Ly \alpha}$ radial profile of our CEN, shown in Fig.~\ref{figures:rpLya}, exhibits a projected distance from the quasar at a SB of $\rm \sim10^{-18} erg~s^{-1}~cm^{-2}~arcsec^{-2}$  (i.e. $\sim 50$~kpc), which is similar to the one inferred from the average profiles from B16 and AB19. This behaviour seems to be independent from the luminosity of the quasar (AB19). However, notice that the SB at this distance coincides with the level of the $2\sigma$ Poisson noise of the image and therefore deeper observations are needed in order to verify this claim.

\begin{figure}
   \begin{center}
   \includegraphics[width=0.48\textwidth]{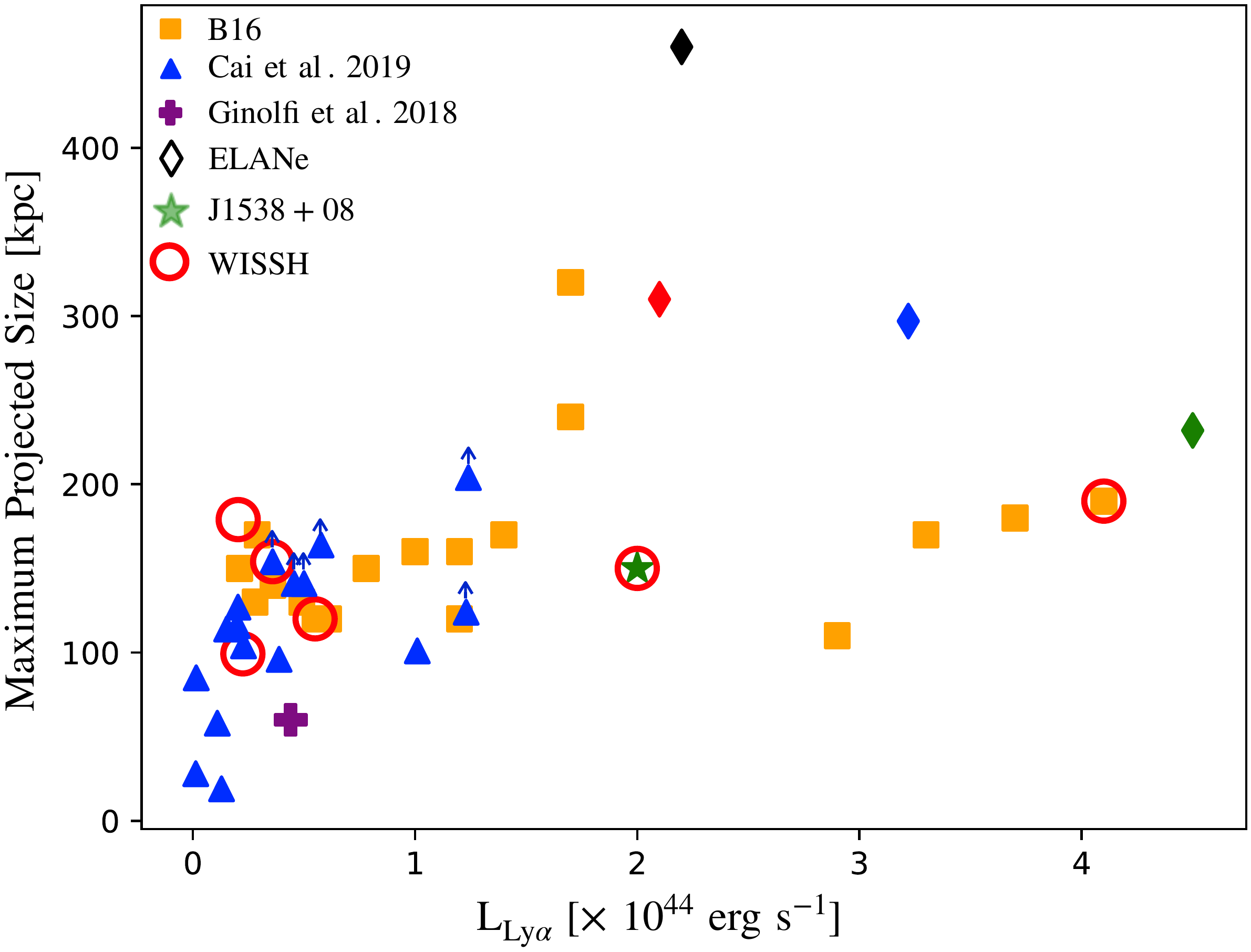}
   \caption{Maximum projected size versus luminosity of the Ly$\alpha$-CEN around J1538+08 (green star) and around quasars from the B16 (orange squares), \cite{Cai19} (blue triangles) and \cite{Ginolfi18} (purple cross) samples. The values for the ELANe (diamond symbols) detected by \cite{Cantalupo14} (black), \cite{Hennawi15} (red), \cite{Cai17} (green) and \cite{Battaia18b} (blue) are also included. WISSH quasars are marked with an additional red circle. We also report the two Ly$\alpha$-CEN around the WISSH quasars from the AB19 sample, i.e. those with the lower luminosities. For these two Ly$\alpha$-CEN the maximum distance from the quasar spanned by the Ly$\alpha$ emission within the 2$\sigma$ isophote multiplied by a factor of two is shown since AB19 do not report the values of the maximum projected size.}
   \label{image:LS}
   \end{center}
\end{figure}

The luminosity of our Ly$\alpha$-CEN is one of the highest measured so far (see Fig.~\ref{image:LS}) and is comparable to the luminosity of known ELANe \citep[e.g,][]{Cantalupo14,Hennawi15,Cai17b,Battaia18b}. 

Interestingly, some WISSH quasars are included in the B16 (J0124+00, J1621$-$00), AB19 (J0125$-$10, J0947+14) and \cite{Cai19} (J2123$-$00) samples. The size and luminosity of the Ly$\alpha$-CEN detected around them appear to be heterogeneous (see Fig.~\ref{image:LS}), suggesting that the properties of these CEN have no simple dependence on the similar, large quasar radiative output (i.e. $\rm L_{bol}$>$\rm 10^{47}~erg~s^{-1}$).

We also checked for possible dependencies on the RQQs bolometric and 2500~\AA~ luminosities (derived from Spectral Energy Distribution modelling; \citealt{WISSHII}; Duras et al. in prep.) to verify if the nuclear radiative output could affect the properties of the Ly$\alpha$-CEN around the WISSH quasars considered here. However, no significant trend with luminosity was observed, confirming the result reported by AB19 for a larger sample of quasars.

\subsection{Kinematics of the Ly$\alpha$- and CIV-CEN of J1538+08}

From the velocity maps of the Ly$\alpha$- and CIV-CEN (Figs.~\ref{image:Lyavelocity} and \ref{image:CIVvelocity}) no coherent kinematic structures possibly hinting at bulk motions and rotations were detected.  
However, the velocity map of the Ly$\alpha$-CEN exhibits a large fraction ($\rm 79 _{-10} ^{+9} \%$)\footnote{Errors derive from the uncertainties to the redshift of both the quasar and nebula.} of pixels with SNR$\geq$3 with negative velocities relative to $\rm z_{CEN}$. This can be justified by the presence of the additional blueshifted component in the Ly$\alpha$ line profile (see Sect.~\ref{sec:skew} and Fig.~\ref{figure:Lyal}). For the CIV-CEN, where we did not detect any significant additional component, the velocity map shows a roughly equal number of pixels with negative and positive velocities. Interestingly, adopting $\rm z_{QSO}$ as reference redshift for the velocity maps of the Ly$\alpha$- and CIV-CEN, we found $\sim$100$\%$ of the SNR$\geq$3 pixels to have negative velocities.  Furthermore, our Ly$\alpha$-CEN shows a peak with a negative velocity of -~438 $\pm$ 267$~\rm km~s^{-1}$. This can be an indication that our $\rm z_{QSO}$ estimated from the H$\beta$ may suffer from systematic uncertainties. Indeed, for a sample of 849 quasars, \cite{Shen16} found that the determination of the quasar redshift from H$\beta$ is subject to uncertainties as large as $\sim$400 $\rm km~s^{-1}$. This can compensate the reported negative velocity offsets of the Ly$\alpha$ peak and the velocity distribution of the SNR$\geq$3 pixels at levels of $\sim$1-2$\sigma$.


Regarding the velocity dispersion, we measured a spatially averaged value $\bar{\sigma _{\rm v}}$= 733 $\pm$ 85 $\rm km~s^{-1}$ from the dispersion map (see Table~\ref{table:NebulaProp}). This is a factor of two larger than those measured around the quasars of B16 and AB19 samples, which span the range $\sigma _{\rm v} \approx[200,400]~\rm km~s^{-1}$. 
This value is more similar to the reported values for Ly$\alpha$-CEN around high-z radio galaxies in the regions affected by radio jets \citep[$\sigma _{\rm v} \gtrsim600~\rm km~s^{-1}$;][]{vanOjik97,Villar03,Humphrey06,Silva18}.

Interestingly, the dispersion map of the Ly$\alpha$-CEN around J1538+08 reports several discrete and compact regions with very high $\sigma _{\rm v}\approx1000~\rm km~s^{-1}$ at an average distance from the quasar of 20-30~kpc (see Fig.~\ref{image:dispersions}). 
These high $\rm \sigma _v$ regions could be due to the turbulence generated by continuum or line-emitting active sources at $\rm z_{CEN}$, possibly injecting energy into the CGM. However, no continuum sources in the MUSE FOV are associated to high $\rm \sigma _v$ values within the Ly$\alpha$-CEN. Similarly to AB19, we used \texttt{CubExtractor} to search for Ly$\alpha$ emitters by setting $\rm SNR_{th}>5$ and $N_{min}^{vox}=50$. We did not detect any source within the CEN. 
We only found a source with a flux of $\rm 3.5 \times 10^{-16}~erg~s^{-1}~cm^{-2}$, at the redshift of the CEN (z$\approx$3.560), outside its boundaries and at a projected distance of $\sim$250 kpc from the quasar (RA = 15:38:28.7, Dec = +8:55:00.80), which very likely cannot affect the nebular emission.

\subsection{Metals in CGM} \label{sec:metcivheii}

The CIV-CEN revealed around J1538+08 is one of the few extended (75 kpc) CIV-emitting regions detected at high significance ($5\sigma$)\footnote{This value represents the integrated SNR computed for the extended CIV emission line.} and spatially mapped around a RQQ. This provides a clear indication of a metal enriched medium (i.e. non-pristine) around this quasar.   
The luminosity of this CEN ($\rm L _{CIV} \simeq 10^{43} \rm erg~s^{-1}$) is comparable to those measured for CIV-CEN detected around radio galaxies at similar redshift \citep{Villar07}. 
However, its morphology shows a marked asymmetry relative to the central quasar position. This is at odds with symmetrically distributed CEN reported around radio galaxies \citep[e.g.][]{Villar07b,Silva18}.
Interestingly the ratio between the SB radial profiles of the CIV and Ly$\alpha$ reported in Fig.~\ref{figure:radproCIV} (i.e. in the southern wedge containing the SB Ly$\alpha$ peak and the CIV-CEN) is a factor of $\sim$0.1 and is constant up to $\rm \sim25$~kpc (i.e. the distance where the Poisson noise starts to dominate the CIV-CEN radial profile). 
This evidence and the spatial co-location of the SB peaks of the two CEN suggest that they could be tracing the same gas and, hence, triggered by the same mechanism. 
Deeper observations would hopefully reveal fainter CIV emission with the same size and morphology of the Ly$\alpha$-CEN in J1538+08.

\subsection{Powering mechanisms for the Ly$\alpha$ CEN}

The possible powering mechanisms of the Ly$\alpha$-CEN include quasar photoionization (i.e. fluorescence), collisional excitation (i.e. cooling) and shocks. In case of fluorescence, assuming that the quasar is surrounded by cold and spherical clouds, we can estimate the CEN SB in two extreme regimes: optically thin (with column density $\rm N_H \ll 10^{17.2} cm^{-2}$) and optically thick \citep[$\rm N_H \gg 10^{17.2} cm^{-2}$;][]{Hennawi13}.
We obtain $\rm SB_{Ly \alpha} ^{thick} \simeq 2.8 \times 10^{-15}~erg~s~cm^{-2}~arcsec^{-2}$ and $\rm SB_{Ly \alpha} ^{thin} \simeq 2.2 \times 10^{-18}~erg~s~cm^{-2}~arcsec^{-2}$, for optically thick\footnote{The luminosity at the Lyman edge, required by the formula (15) by \cite{Hennawi13}, has been estimated by the $\rm L_{1450}$ (see Duras et al. in prep.) as detailed in  \cite{Lusso15} and \cite{Farina19}. Moreover, we adopted as CEN radius 40 kpc (see Fig.~\ref{figures:rpLya}) and a covering factor $\rm f_c=0.5$} and optically thin\footnote{We used the formula (10) in \cite{Hennawi13} and adopted the following fiducial values: $\rm N_H = 10^{20}~cm^{-2}$, density $\rm n_H = 0.1~cm^{-3}$ and $\rm f_c = 0.5$.} gas, respectively.  In our case, we found that $\rm SB_{Ly \alpha}$ for regions with SNR$\geq$3 is $\rm \simeq 2.3 \times 10^{-17}~erg~s~cm^{-2}~arcsec^{-2}$.
This value is more compatible to $\rm SB_{Ly \alpha} ^{thin}$ than to $\rm SB_{Ly \alpha} ^{thick}$, supporting the presence of an optically thin medium as already reported in previous works (e.g. AB19, \citealt{Cai19}).

The CIV/Ly$\alpha$ and HeII/Ly$\alpha$ ratios could be used to understand if shocks or collisional excitation are viable powering mechanisms for the observed Ly$\alpha$-CEN \citep[but see][]{Cantalupo19}. The detection of the CIV-CEN allows us to rule out cooling due to gravitational accretion as the powering mechanism of this CEN. However, it is notoriously difficult to disentangle photoionization models from shock models in a HeII/Ly$\alpha$ vs CIV/Ly$\alpha$ diagram \citep{Battaia15b}. Nevertheless, photoionization models usually do not predict lower levels of HeII/Ly$\alpha$ with respect to CIV/Ly$\alpha$ \citep[e.g.,][]{Humphrey19,Cantalupo19}. On the contrary they predict ratios of the same order for the two transitions. This is because these two emission lines have similar ionization energies 64.5 eV for CIV and 54.4 eV (4 Ryd) for HeII.
On the other hand, shock models \citep{Allen08} do show cases in which the HeII/Ly$\alpha$ ratio is lower than the CIV/Ly$\alpha$ ratio. 
Indeed, according to shock+precursors models presented by \cite{Allen08} and \cite{Battaia15a} for the origin of extended nebular emission, the values of CIV/Ly$\alpha$ and HeII/Ly$\alpha$ inferred from our analysis are consistent with a shock propagating at 200-300 $\rm km~s^{-1}$ in a $\rm n_H \sim 10-100~cm^{-3}$ gas or with a faster >1000 $\rm km~s^{-1}$ shock in a denser gas ($\rm n_H>100~cm^{-3}$). If we assume that the $\rm v_{shift} =  1520~km~s^{-1}$ is the velocity of the shock, then the emitting gas is required to be at high densities ($\rm n_H > 100 ~cm^{-3}$).

The aforementioned photoionization and shock models do not include a contribution from resonant scattering of Ly$\alpha$ photons from the quasar. In the case such contribution is important, the CIV/Ly$\alpha$ and HeII/Ly$\alpha$ ratios predicted by those models needs to be corrected and shifted to lower values. However, it has been shown that the efficient diffusion in velocity space allows the Ly$\alpha$ resonantly scattered photons produced by the quasar itself to escape the system at very small scales, $<10$~kpc \citep{Dijkstra06}. Therefore, this should result in a negligible contribution of scattered Ly$\alpha$ emission on scales $>10$~kpc \citep[e.g,][]{Cantalupo14}.
A firm characterization of the Ly$\alpha$ resonant scattering contribution requires a full radiative transfer calculation on a three dimensional gas distribution representing this system. This approach is beyond the scope of this work.


\subsection{Evidence of outflowing gas in the CGM}

The blueshifted broad component in the Ly$\alpha$ spectrum of the CEN provides a tantalizing evidence for the presence of outflowing gas reaching a projected distance of 20-30~kpc from the quasar, i.e. into the CGM surrounding J1538+08. 

J1538+08 also exhibits a host-galaxy scale [OIII] outflow with mass outflow rate $\rm \sim 530~ M_{\odot} ~yr^{-1}$, maximum velocity $\rm \sim 2900 ~km~s^{-1}$ and kinetic power of $\rm \dot{E} _{kin} = 1.4  \times 10^{45}~erg~s^{-1}$ \citep[see][]{Vietri18}.
In an energy conserving scenario in which the large-scale outflow reported in this paper is the later stage of the AGN-driven [OIII] one \citep[e.g.][]{WISSHI,Vietri18}, by assuming an expansion at $\rm v_{out} = v_{shift} \sim 1500~km~s^{-1}$, we estimated a mass outflow rate of $\rm \dot{M}_{out} \simeq 300 ~M_{\odot}~yr^{-1}$, comparable to the [OIII] one. 
Notice that, assuming $\rm v_{out} = 300~km~s^{-1}$ as suggested by the CIV/Ly$\alpha$ and HeII/Ly$\alpha$ values in a shock+precursor models (see Sect.~\ref{sec:metcivheii}) we would obtain a $\rm \dot{M}_{out}$ larger by a factor of 25, but still comparable with the typical $\rm \dot{M}_{out}$ reported for [OIII] outflows in the WISSH sample \citep{WISSHI}.
By assuming a constant velocity of 1500$\rm~km~s^{-1}$ (see Tab.~\ref{tab:fit}), we estimated that an outflow at host galaxy scales \citep[i.e. $\sim$5-10 kpc ][]{WISSHI} would take $\sim$7-13 Myr to reach the distance of our CGM outflow (i.e. $\sim$30 kpc). Adopting a velocity of $\sim $300$\rm ~km~s^{-1}$, we would obtain an upper limit to the outflow time a factor of 5 larger than the previous estimate.

The sudden increase of the velocity dispersion of the Ly$\alpha$-CEN to $\sigma _{\rm v} \sim$900$\rm~km~s^{-1}$ at a distance of $\sim$20 kpc (see Fig.~\ref{image:RPdispvel}) could be partially due to the presence of the outflow component, hence to an incorrect parameterization of the line profile with a single Gaussian component. Unfortunately the SNR of our observations does not allow us to recompute the radial profile of the $v$ and $\sigma _{\rm v}$ of the Ly$\alpha$ line using a two-component Gaussian modelling, because the SNR of the broad Ly$\alpha$ line in each bin is too low.
Deeper MUSE observations are therefore needed to verify the presence of a broad and blueshifted component over the entire Ly$\alpha$-CEN and assess if the CIV-CEN also shares the same kinematics.

Given the resonant nature of the Ly$\alpha$ transition, complex radiative transfer models accounting for the  velocity field, density structure of the CEN and the geometry of the radiation field are needed in order to properly interpret the properties derived by the Ly$\alpha$ line profile fitting \citep[e.g.][]{Cantalupo05,Gronke16}. 
It is worth noting that the presence of an expanding shell surrounding the quasar, assuming spherical symmetry and isotropic geometry of the photoionizing radiation field, may produce an asymmetric double peak profile in which the blue-ward component is strongly suppressed \citep{Verhamme06,Laursen09,Steidel10,Chung19}. This would provide support to the hypothesis of an outflowing CEN whose two components of the Ly$\alpha$ profile are associated with the same expanding gas. 
Notice that in this simple model (i.e. symmetric, homogeneous and isotropic gas distribution) the main peak component is always on the red side compared with the systemic redshift of the CEN. 
However, assuming that $z_{\rm QSO}\equiv z_{\rm CEN}$, our Ly$\alpha$-CEN does not show such a red line component (see Fig.\ref{image:FitLya}). 
To test the nature of this putative outflow in the context of an asymmetric, inhomogeneous gas distribution, MUSE observations with higher SNR are needed along with a more accurate determination of $z_{\rm QSO}$ (e.g. from the CO line) and radiative transfer models.

\section{Summary and conclusions}\label{conclusions}
In this paper we have presented a VLT/MUSE investigation on the CGM around J1538+08, a z$\approx$3.6, broad-line, RQQ belonging to the WISSH quasars sample \citep{WISSHI}. The main results can be summarized as follows:\\

\begin{itemize}
   \item we discovered a CGM Emission Nebula (CEN) detected in Ly$\alpha$ of $\sim$150~kpc surrounding J1538+08, one among the most luminous Ly$\alpha$-CEN ($\rm \sim 2 \times 10^{44}\rm~erg~s^{-1} $) reported so far \citep{Borisova16,Battaia19,Farina19}. Our nebula appears roughly symmetric on large scales (several tens of kpc) and exhibits a bright SB peak located at $\sim10-15$~kpc southward of the quasar;

   \item we obtained one of the first 2D-mapping of a significantly detected ($\sim 5\sigma$), extended ($\sim$75~kpc) CIV-CEN around a RQQ. Given its spatial coincidence with the Ly$\alpha$ SB peak and a similar SB profile, it is very likely associated with the Ly$\alpha$-CEN;

   \item we found no significant velocity pattern in the kinematics of the Ly$\alpha$-CEN. Remarkably, the average velocity dispersion $\bar{\sigma _{\rm v}} \simeq 700~ \rm km~s^{-1}$ is higher than the typical values measured in RQQs and much more similar to the dispersion observed for Ly$\alpha$-CEN around high-redshift radio galaxies  \citep{vanOjik97,Villar03,Humphrey06,Silva18} and outflow-dominated systems \citep{Ginolfi18}.

   \item We obtained one of the first 2D characterization via IFU spectroscopy of an ionized outflow at CGM scales ($\gg$10 kpc) around a RQQ, by performing the spectral analysis of an extended region with negative skewness value.
   
   \begin{itemize}
       \item Specifically, the analysis of the skewness map of the Ly$\alpha$-CEN reveals a region within 30 kpc and to the south of the quasar, in which the skewness is negative (see Fig.~\ref{image:skew}) and the Ly$\alpha$ emission profile is significantly asymmetric with a blue tail. This region roughly overlaps with the SB peak of the Ly$\alpha$-CEN and includes the CIV-CEN;
   
       \item the Ly$\alpha$ spectrum extracted from the region showing negative skewness is well modelled with two Gaussian components (Fig.\ref{image:FitLya}). This fit resulted into a systemic narrow  ($\sigma _{\rm v}\sim500~\rm km~s^{-1}$) component and a broader ($\sigma _{\rm v} \approx$ 1200$~\rm km~s^{-1}$) one. The latter is blueshifted by $v_{\rm shift} \simeq 1500~\rm km~s^{-1}$ and is indicative of outflowing gas on CGM-scales.

   \end{itemize}
   
   
\end{itemize}

All the reported results clearly indicate the presence of a metal-enriched (i.e. non pristine) gas with kinematic features consistent with an outflowing gas component at scales of tens of kpc.

Both deeper spatially resolved spectroscopic observations of the CGM around this hyper-luminous quasar and dedicated radiative transfer modellings are necessary in order to confirm and refine this scenario.
Specifically, they are needed to accurately characterize the CEN and outflow physical properties and understand the role of the outflow in transporting metals in the CGM.


\begin{acknowledgements}
  We thank Laura Pentericci, Fabrizio Nicastro, Emanuele Giallongo and Marco Stangalini for useful discussions.
  We thank Gabriele Pezzulli for providing us with the surface brightness radial profile of J0124+00.
  LZ, EP acknowledge financial support under ASI/INAF contract 2017-14-H.0.
  FF, EP, AB and CF acknowledge financial support from PRIN-INAF-2016 FORECAST.
  This research has made use of the NASA/IPAC Extragalactic Database (NED), which is operated by the Jet Propulsion Laboratory, California Institute of Technology, under contract with the National Aeronautics and Space Administration.
  SC gratefully acknowledges support from Swiss National Science Foundation grant PP00P2$\_$163824.

\end{acknowledgements}

\bibliographystyle{aa} 
\bibliography{andrea.bib}


\appendix 

\section{Stellar contamination} \label{Saturstar}

\begin{figure}
   \begin{center}
   \includegraphics[width=0.48\textwidth]{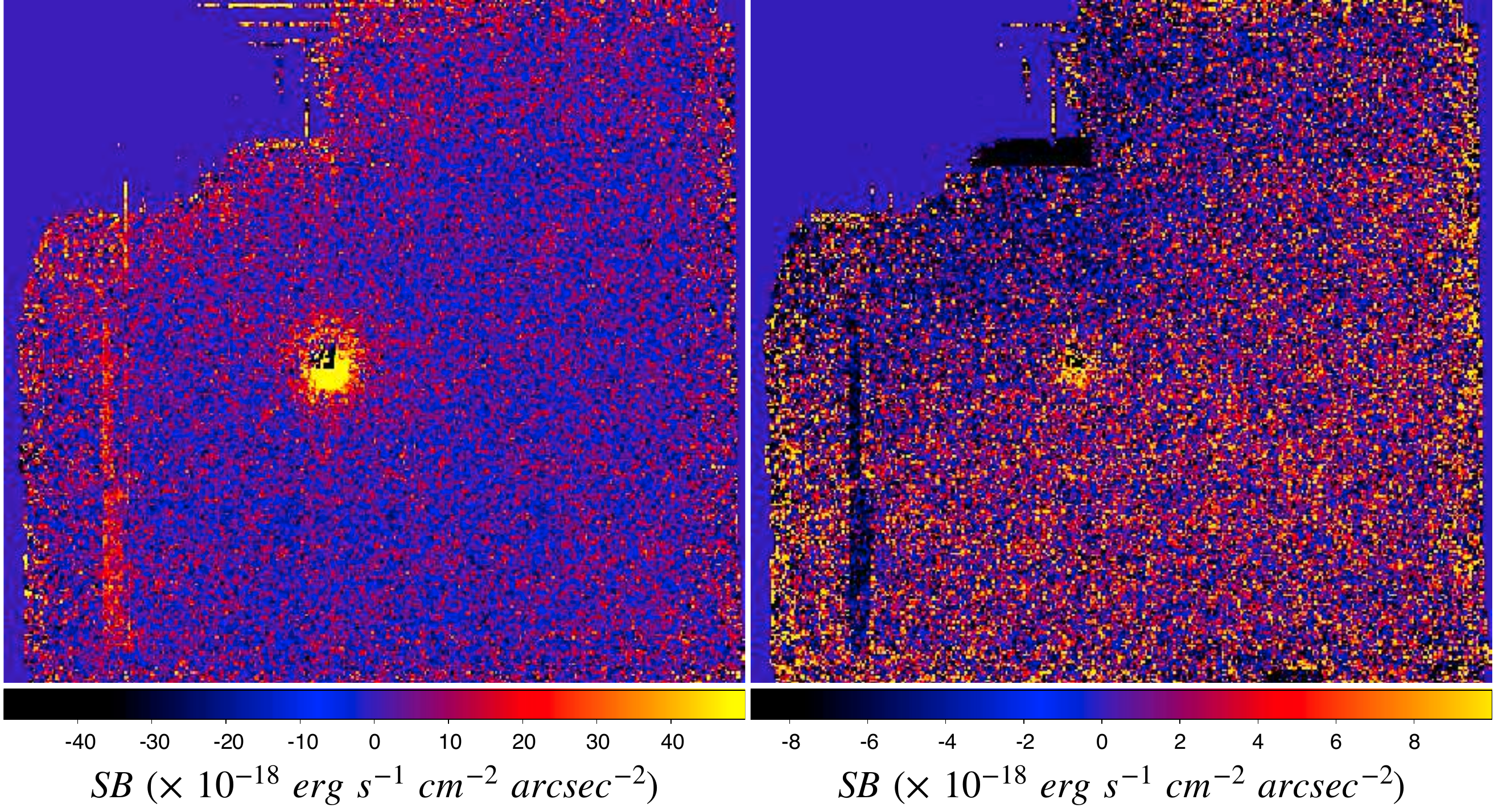}
   \caption{Images obtained by collapsing the final PSF- and continuum- subtracted datacubes in the spectral regions of Ly$\alpha$- (left panel) and CIV-CEN (right panel), respectively. The homogeneous blue area on the upper-left corner represents the mask applied to the exposures during the data reduction in order to exclude the saturated stellar flux in the FOV.}
   \label{image:WIstarsatura}
   \end{center}
\end{figure}

Fig.~\ref{image:WIstarsatura} reports the image of the MUSE FOV obtained by collapsing the spectral region of the Ly$\alpha$- (left) and CIV- (right) CEN in the PSF- and continuum- subtracted datacube. The upper-left corner in each image represents the region contaminated by the luminous star, which has been masked during the data reduction.
A visual inspection reveals no residual contamination at the CEN position. We performed a more quantitive estimation by measuring the average flux in concentric annular regions centered on the star. We found no significant stellar contamination on the nebular emissions.
The average contamination was estimated in concentric annuli with respect to the center of the star, by excluding the nebular region. We estimated a stellar contamination of the order of $\sim$11$\%$, $\sim$6$\%$ and $\sim$6$\%$ in north and south Ly$\alpha$-CEN regions and over the entire CIV-CEN, respectively.


\section{Blue tail} \label{sec:BTwing}

\begin{figure*}
   \begin{center}
   \includegraphics[width=0.95\textwidth]{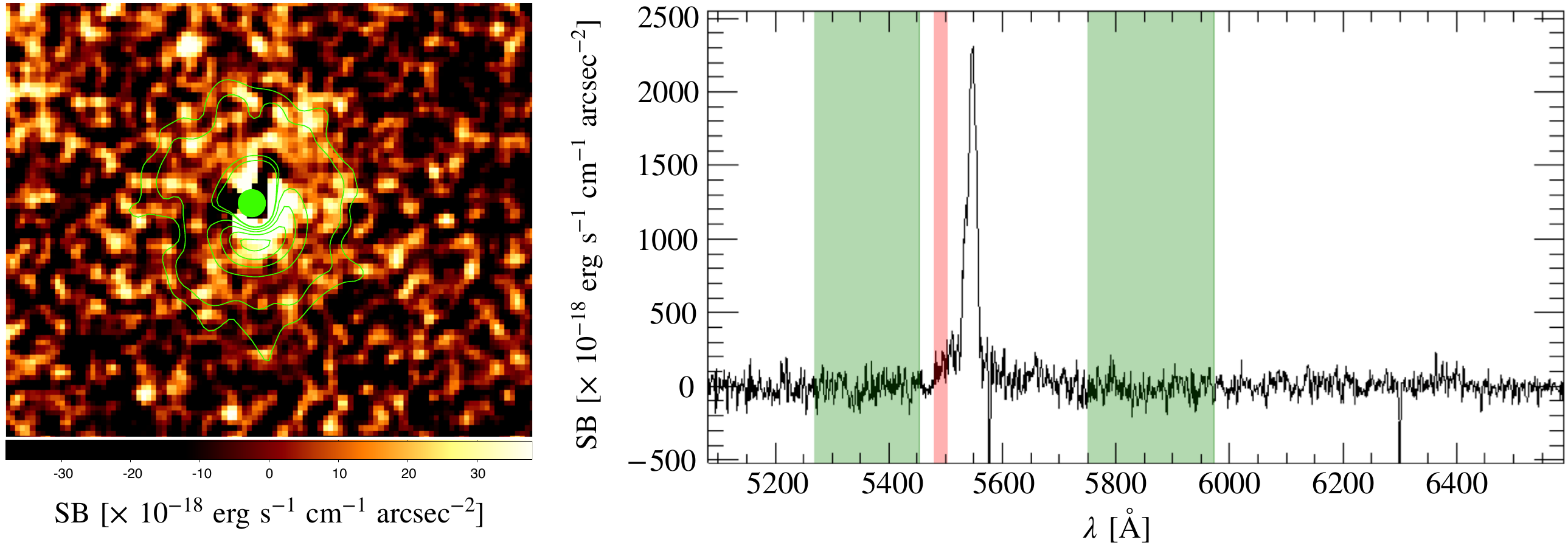}
   \caption{Left panel: pseudo-NB image obtained by collapsing the spectral region of the blue tail (see red area in the right panel) and by subtracting the continuum estimated in spectral regions without line features (see green areas in the right panel). Right Panel: spectrum which is extracted from the PSF- and continuum-subtracted datacube, by selecting only the spaxels belonging to the CEN with SNR>5. The red and green areas mark the spectral region collapsed and subtracted, respectively, to obtain the pseudo-NB in the left panel. }
   \label{image:WWW}
   \end{center}
\end{figure*}

In this Appendix we report on several tests we performed in order to check for possible contamination of the weak CEN emission from the bright quasar emission lines residuals after the Ly$\alpha$ subtraction.
Left panel of Fig.~\ref{image:WWW} shows the SB emission of more blueshifted portion of the blue tail (i.e. highlighted in red in the right panel). This was obtained by collapsing the spectral region including the wing of the blue tail, in order to avoid any possible contamination of the narrow Ly$\alpha$ component. To remove possible residuals at each pixels we applied a subtraction of the continuum, measured from the spectral region highlighted in green in Fig.~\ref{image:WWW}.
The final pseudo-NB image exhibits an asymmetric shape with respect to the quasar position and it is spatially associated with the peak of the SB of the Ly$\alpha$-CEN. In case of residual AGN contribution, we would have expected a symmetric emission around the quasar. This provides an indication that the blue tail is not the result of emission line contamination from the quasar Ly$\alpha$.

We found that the average intensity of the PSF subtraction residuals are 3 times lower than the intensity of the blue tail, which is, moreover, spatially coincident with a high SNR region.

We further check the quasar emission line contamination by inspecting the region of the bright quasar SiIV emission line. We measured the PSF-subtracted radial profile of the slope of the blue tail at the wavelength of the SiIV line. We found a flat radial profile with an average value around zero. This provides a further indication that the quasar emission does not contribute to the blue tail emission and hence, that this is a genuine and intrinsic emission of the CEN.

\end{document}